\documentstyle[12pt,epsf]{article}
\renewcommand{\baselinestretch}{1.0}

\newcommand{\be}{\begin{equation}}
\newcommand{\ee}{\end{equation}}
\begin{document}
\topmargin 0pt
\oddsidemargin=-0.4truecm
\evensidemargin=-0.4truecm
\renewcommand{\thefootnote}{\fnsymbol{footnote}}

\newpage
\setcounter{page}{1}
\begin{titlepage}     
\vspace*{-2.0cm}
\begin{flushright}
\vspace*{-0.2cm}
\end{flushright}
\vspace*{0.5cm}

\begin{center}
{\Large \bf Global Analysis with SNO:\\ Toward the Solution of the Solar
Neutrino Problem}
\vspace{0.5cm}

{P. I. Krastev$^{1}$ and  A. Yu. Smirnov$^{2,3}$\\
\vspace*{0.2cm}
{\em (1) Philips Analytical, Natick, MA 01760, USA }\\
{\em (2) The Abdus Salam International Centre for Theoretical Physics,  
I-34100 Trieste, Italy }\\
{\em (3) Institute for Nuclear Research of Russian Academy 
of Sciences, Moscow 117312, Russia}

}
\end{center}

\vglue 0.8truecm
\begin{abstract}

We perform a global analysis of the latest solar neutrino data
including  the Solar Neutrino
Observatory (SNO) result on the $CC-$event rate. This result
further favors the LMA MSW solution of the solar neutrino problem. 
The best fit values of parameters we find are: 
$\Delta m^2 = (4.8 - 5.0) \cdot 10^{-5}$
eV$^2$, $\tan^2 \theta = 0.35 - 0.38$, $f_B = 1.08 - 1.12$, and
$f_{hep} = 1 - 4$, where $f_B$ and $f_{hep}$ are the boron and $hep-$
neutrino fluxes in units of the corresponding fluxes in the Standard
Solar Model (SSM).  With respect to this best fit point the LOW MSW  
solution is accepted at 90 \% C.L..   
The Vacuum oscillation solution (VAC) with 
$\Delta m^2 = 1.4 \cdot 10^{-10}$ eV$^2$,  gives good fit of the data 
provided that the boron neutrino 
flux is substantially smaller than the SSM flux ($f_B \sim 0.5$). 
The SMA solution is accepted  at about $3 \sigma$
level. We find that vacuum oscillations to sterile neutrino,  
VAC(sterile), with $f_B \sim 0.5$ also give rather 
good global fit of the data. 
All other sterile neutrino solutions are strongly disfavored.  
We check the quality of the fit by constructing the 
{\it pull-off} diagrams of observables for the global solutions. 
Maximal mixing is allowed at $3\sigma$ level in the LMA region and at 
95 \% C.L. in the LOW region.  Predictions for the day-night asymmetry, 
spectrum distortion and  ratio of the neutral to charged curent event 
rates, [NC]/[CC], at SNO  
are calculated.  In the best fit points of the global solutions we find:
$A_{DN}^{CC} \approx (7 - 8)\%$ for LMA, $\sim 3 \%$ for
LOW, and $(2 - 3) \%$ for SMA.  In the LMA region the asymmetry can be
as large as 15 - 20 \%.  Observation of $A_{DN}^{CC} > 5 \%$ will
further favor the LMA solution.  It will be difficult to see the
distortion of the spectrum expected for LMA 
as well as LOW solutions. However, future SNO 
spectral data can significantly affect the VAC and SMA solutions.
We  present  expected values of the BOREXINO event rate     
for global solutions.  

\end{abstract}
\centerline{Pacs numbers: 14.60.Lm 14.60.Pq 95.85.Ry 26.65.+t} 
\end{titlepage}
\renewcommand{\thefootnote}{\arabic{footnote}}
\setcounter{footnote}{0}
\renewcommand{\baselinestretch}{0.9}

\section{Introduction}

With the SNO result \cite{sno,sno1} on the $CC-$rate, we now, for the
first time ever, have more than $3\sigma$ evidence of flavor
conversion of solar neutrinos.  ``Smoking guns'' have indeed started
to smoke.  The statement, made first in the original publication by
the SNO collaboration, is based on difference of the boron neutrino
fluxes determined from the charged current ($CC-$) event rate in the
SNO detector, and the $\nu e-$scattering event rate obtained by the
SuperKamiokande (SK) \cite{SK} collaboration (and confirmed, albeit
with smaller statistical significance, by SNO). Somewhat
paradoxically, the absence of significant distortion of the boron
neutrino spectrum at SuperKamiokande adds to the strength of this
conclusion.

In general, the difference between the signals in the SNO and SK
detectors can be due to:

1) appearance of the  $\nu_{\mu}, \nu_{\tau}-$ flux, or/and 

2) distortion of the neutrino energy spectrum.  If the suppression of
the boron neutrino flux (due to some neutrino transformations)
increases with neutrino energy, then the higher sensitivity to higher 
energies of the $CC-$reaction in SNO as compared with $\nu-e$
scattering in SK  would explain the lower rate in SNO.

In fact, both reasons imply neutrino conversion.  However, an absence
of strong distortion of the boron neutrino spectrum, as found by
SuperKamiokande and independently confirmed by SNO, leads to the
conclusion that the main reason for the difference is the appearance
of the $\nu_{\mu}, \nu_{\tau} -$ flux.

From the SNO result  and its comparison with the SK data 
one can immediately draw several conclusions:

\begin{itemize}

\item
Strong deficit of the $\nu_e-$ flux with respect to  the 
Standard Solar Model (SSM) predictions is found. 

\item
There is a strong evidence that  $\nu_e$ from the sun 
are converted to $\nu_{\mu}, \nu_{\tau}$.  

\item
There is no astrophysical solution of the solar neutrino problem. 
(For recent quantitative analysis see~\cite{astro}). 

\item
Solutions of the solar neutrino problem based on pure 
active - sterile conversion, $\nu_e \rightarrow \nu_s$,
are strongly disfavored. 

\item
More than half of the original $\nu_e-$flux is transformed to
neutrinos of different type $\nu_{\mu}$, $\nu_{\tau}$, and possibly
$\nu_s$, that is, the $\nu_e-$survival probability
\be
P < 1/2. 
\label{half}
\ee
In fact, in assumption of pure active transition one gets $P = 0.334
\pm 0.22$ \cite{sno1} which is just $ \sim 1 \sigma$ below 1/2.
Clearly, if the sterile neutrino component is present in the flux, the
survival probability is even smaller.  The inequality (\ref{half}), if
confirmed, will have crucial implications for further experimental
developments in the field, as well as for fundamental theory.

\end{itemize}

We know now with a high confidence  that electron neutrinos
produced in the center of the Sun undergo flavor conversion.  However
the specific mechanism of the conversion has not yet been identified.
As we will see, only some extreme possibilities are excluded by adding
the SNO result to the previously available solar neutrino data. A
number of solutions still exist.

The SNO result changes the status of specific solutions to the solar
neutrino problem. The changes can be immediately seen by comparing
predictions for the $CC-$ event rate from global solutions found in
the pre-SNO analysis \cite{bks01,sno-sk} with the SNO result.  There
are four solutions, three active and one sterile, which give 
predictions close to the SNO result:
\be
R^{SNO}_{CC} \equiv \frac{SNO}{SSM} =  0.347 \pm 0.029. 
\ee

1). The Large Mixing Angle MSW solution (LMA): the 
$3\sigma$ predictions interval,  
$R_{CC} = 0.20 - 0.41$, covers the SNO result ($R_{CC} \equiv$ [CC] from
\cite{bks01}). 
The best fit point from the pre-SNO analysis, $R_{CC} = 0.31$, is
slightly ($\sim 1
\sigma$) below  
the central value given by SNO. Therefore, SNO shifts the region of
the LMA solution and the best fit point to larger values of mixing
angles which correspond to larger survival probability.

2). The low $\Delta m^2$ (LOW) solution:  the interval of predictions,
$R_{CC} = 0.36 - 0.42$, is above the SNO result.  In the best fit
point $R_{CC}$ is $2 \sigma$ (experimental) higher the central SNO
value. Therefore this solution is somewhat less favored, and SNO tends
to shift the allowed region to smaller values of $\theta$ which correspond
to
smaller survival probability.

3). The vacuum oscillation solutions (VAC): the expected  
interval,  $R_{CC} = 0.33 - 0.42$, 
is also covering the SNO range, and the best fit value, $R_{CC} =
0.38$, is just $1 \sigma$ above the central SNO result. Therefore, SNO
improves the status of this solution.

4). Vacuum oscillations to sterile neutrinos, 
VAC(sterile):  the predicted
interval, $R_{CC} = 0.36 - 0.41$, with
the best fit point 0.39, is only slightly higher than for the
VAC(active) case.  Here low rate predicted for SNO is due to difference of
the thresholds in SNO and SK  and to the steep  
increase of  the suppression with neutrino  energy.  
This solution survives the first SNO
result.

Other solutions look less favorable in the light of the SNO result. In
particular, the small mixing angle MSW (SMA) solution interval,
$R_{CC} = 0.37 - 0.50$, (especially its best fit point, 0.46,  from the
pre-SNO analysis) is substantially above the SNO rate. This solution
is further disfavored by SNO.  In fact, SMA predicts opposite energy
dependence of the suppression on the neutrino energy to the one SNO
prefers.

The Just-so$^2$(active), SMA(sterile) and Just-so$^2$(sterile)
solutions predict practically the same rate for the SK and SNO and
therefore are strongly disfavored.

The statements above are  supported by detailed quantitative
analysis of the solar neutrino data we have performed in this paper. 
The results of this analysis are described in the rest of this paper.  
Previously
some of the same conclusions have been reported in
\cite{barger,fogli,valencia,calcutta}; see also related studies
\cite{gouvea,BS,giunti,berezinsky}. 

The  paper is organized as follows: In Section 2 we present
the results of the global analysis of the solar neutrino data.  We
construct the pull-off diagrams of available observables for the found
global solutions. In Section 3 we consider specifics of global
solutions and their implications. We evaluate the quality of the data
fit in each of the global solutions.  We calculate predictions for the
day-night asymmetry and spectrum distortion in Section 4.  Finally,
prospects for identifying the solution to the solar neutrino problem
are discussed in Section 5. \\

\section{Global analysis and pull-off diagrams}

We describe here the results of the global analysis of the solar
neutrino data.  We find global solutions (sets of oscillation
parameters and solar neutrino fluxes which explain the solar neutrino
data) and determine the goodness of fit (g.o.f.) in the best fit
points.  We perform diagnostics of the global solutions by checking
their stability with respect to variations in the analysis and to
uncertainties in the original solar neutrino fluxes.  To check the
quality of the fit we  construct  the pull-off diagrams for the
observables.

\subsection{Features of the analysis}

We follow the procedure of the analysis developed in our previous
publications \cite{bks98,bks-10,bks-corr,bks01} in collaboration with
John Bahcall.  We also describe some additional features, which should
be taken into account when comparing our results with those obtained
by other groups.

In our global analysis we use: (i) the $Ar-$production rate from
the Homestake experiment \cite{Cl}, (ii) the $Ge-$production rate from
SAGE \cite{sage}, (iii) the combined $Ge-$production rate from GALLEX
and GNO \cite{gallex,gno}, (iv) the $CC-$event rate measured by SNO
\cite{sno}, (v) the  Day and Night energy spectra 
measured by SuperKamiokande \cite{SK}. 

Following the procedure outlined in \cite{bks01} we do not include the
total rate of events in the SK detector, which is not independent from
the spectral data.  In our standard global analysis, we use 4 rates
and 38 spectral data, a total of 42 data points. The number of free
parameters and the number of d.o.f. are different in different
analyses and we will specify them later.

The solar neutrino fluxes are taken according to SSM BP2000 \cite{ssm}
with the corrected (due to improved measurement of the solar
luminosity) boron neutrino flux $F^{SSM}_B = 5.05 \cdot 10^{6}$
cm$^{-2}$ c$^{-1}$.  We denote by $f_B$ and $f_{hep}$ the fluxes of
the boron and $hep-$neutrinos measured in units of the BP2000 fluxes.

The analysis of the data is performed in terms of two neutrino mixing
characterized by the mass squared difference $\Delta m^2$ and the
mixing parameter $\tan^2\theta$. 
We consider conversion into  pure active or pure
sterile neutrinos.

We perform the $\chi^2$ test of various oscillation solutions by
calculating

\be
\chi^2_{global} = \chi^2_{rate} +   \chi^2_{spectrum}, 
\label{chi-def}
\ee
where $\chi^2_{rate}$ and $\chi^2_{spectrum}$ are the contributions
from the total rates and from the SuperKamiokande day and night spectra
correspondingly. Each of 
the entries in Eq.(\ref{chi-def}) is a function of the four parameters
($\Delta m^2$, $\tan^2\theta$, $f_B$ and $f_{hep}$). It is an
important feature of our approach \cite{bks01} that at each step of the
minimization process these parameters are kept the same in each of the
two entries on the right hand side of Eq. (\ref{chi-def}).

\subsection{Free flux fit}

We perform the fit to the experimental data treating the boron neutrino
flux, $f_B$, and the $hep-$neutrino flux, $f_{hep}$, as free
parameters.  There are several reasons to consider $f_B$ and $f_{hep}$
as free parameters (for earlier work in this context see \cite{KSm}):

1.  These fluxes have the largest uncertainties in the SSM (see
however, \cite{hepfl}). 

2.  The goal of the solar neutrino studies is to find directly from
the solar neutrino data both the oscillation parameters and the
neutrino fluxes.  The free flux analysis is the   way to achieve
this goal.

3. Comparing the fluxes $f_{hep}$ and $f_B$ found from the free flux
analysis with the SSM fluxes we can estimate the plausibility of the fit
and the reliability of the solutions.  Clearly, strong deviation of the
fluxes for a given solution from the SSM values will indicate certain
problems with either the solution or the SSM predictions.

4. Last, but not least, this fit is the most conservative one
regarding the exclusion of certain scenarios.

\begin{table}
\caption[]{Best-fit values of the parameters $\Delta
m^2$,  $\tan^2\theta$, $f_B$ and $f_{hep}$ from the 
free flux analysis.   The minimum $\chi^2$ and the
corresponding g.o.f. are given in the last two columns. 
The number of degrees of freedom is 38: 
4 rates (Homestake,  SAGE, Gallex/GNO, SNO) + 38 SK spectra points  - 
4 parameters.}
\vskip 0.5cm
\begin{tabular}{l c c c c c c }
\hline 
Solution & $\Delta m^2$/$\rm eV^2$ &    $\tan^2\theta$ & $f_B$ &
$f_{hep}$ & $\chi^2_{min}$ & g.o.f. \\ 
\hline
LMA      & $4.8\times 10^{-5}$ &  0.35   & 1.12 & 4 &   29.2 & 0.85  \\

VAC      & $1.4\times 10^{-10}$ & 0.40 (2.5)  & 0.53 & 6 & 32 & 0.74 \\

LOW      & $1.1\times 10^{-7}$ &   0.66  & 0.88  & 2   & 34.3 & 0.64 \\

SMA      & $6.0\times 10^{-6}$ &  $1.9\times 10^{-3}$ & 1.12  & 4 & 40.9 
& 0.34 \\

Just So$^2$ & $5.5\times 10^{-12}$ & 1.0 & 0.44 & 0 &   45.8 & 0.18 \\

VAC(sterile) & $1.4\times 10^{-10}$ & 0.38 (2.6) & 0.54 & 9 & 35.1 & 0.60 
\\

Just So$^2$(sterile) & $5.5\times 10^{-12}$ & 1.0 & 0.44 & 0 & 46.2 &
0.17 \\

SMA(sterile) & $3.8\times 10^{-6}$ & $ 4.2\times 10^{-4}$ & 0.52 & 0.2 &
48.2 & 0.12\\

LMA(sterile) & $1.0\times 10^{-4}$ & 0.33 & 1.14 & 0 & 49.0 & 0.11  \\  

LOW(sterile) & $2.0\times 10^{-8}$ & 1.05  & 0.83 & 0 & 49.2 & 0.11 \\

\hline
\end{tabular}
\label{Table1}
\end{table}  

Thus, together with $\Delta m^2$ and $\tan^2 \theta$, there are 4
free parameters and therefore 42 - 4 = 38 d.o.f. in the $\chi^2$-fit.
In Table \ref{Table1}  we show the best fit values of the parameters 
$\Delta m^2$,  $\tan^2\theta$, $f_B$, $f_{hep}$ for different solutions 
of the solar neutrino problem. We also give the corresponding 
values of $\chi^2_{min}$ and the goodness of the fit.    
Several remarks are in order. 

The absolute $\chi^2$ minimum: $\chi^2 = 29.2$ is in the LMA region.
Such a low $\chi^2$ for 38 d.o.f. is mainly due to the small spread of
the experimental points in both the day and the night SK spectra. 
Similar situation is realized for the LOW solution. 
The SuperKamiokande day and night spectra can be especially well 
described by solutions which predict a  bump in the survival probability  
at $E = 6 - 8$ MeV  and a dip at $E = 10 - 11$ MeV. This is the case of
VAC solutions (both active and sterile) with large $hep-$flux. It is for
this reason VAC solutions have high goodness of the global fit.       
According to the Table~\ref{Table1} the VAC(sterile) solution is in the
third position after LMA and VAC(active) and its fit is even better than
the one of  LOW solution.  The VAC(sterile) solution has, however, a
number of problems which we will discuss in the next section.

For LMA, LOW and SMA solutions values of $f_B$, agree with
the SSM predictions within $1 \sigma$ theoretical uncertainty 
($\sim 18$ \%).  All VAC solutions and SMA(sterile), as well as
Just-so$^2$ solutions, appear with a boron flux which 
is $3 \sigma$ below the SSM
boron neutrino flux.  For the $hep-$neutrino flux the VAC and SMA
solutions imply significant (factors 4 - 6) deviation from the central
SSM value. The VAC(sterile) requires as large as 9 SSM $hep-$neutrino 
flux.

For VAC solutions the matter effect is  negligible  and in
the Table 1,  two ``symmetric" values of $\tan^2 \theta$:   
$\tan^2 \theta_1 = 1/\tan^2 \theta_2$ 
correspond to mixing in normal and 
dark (in brackets) sides of the parameter space. 

Even within this conservative analysis the LMA(sterile) and
LOW(sterile) solutions give a very bad fit and in what follows we will
not discuss them. 

Just-so$^2$(active) and Just-so$^2$(sterile) give very similar description
of the data. So, we will  present results for one of them.

\begin{figure}[ht]
\centering\leavevmode
\epsfxsize=1.0\hsize
\epsfbox{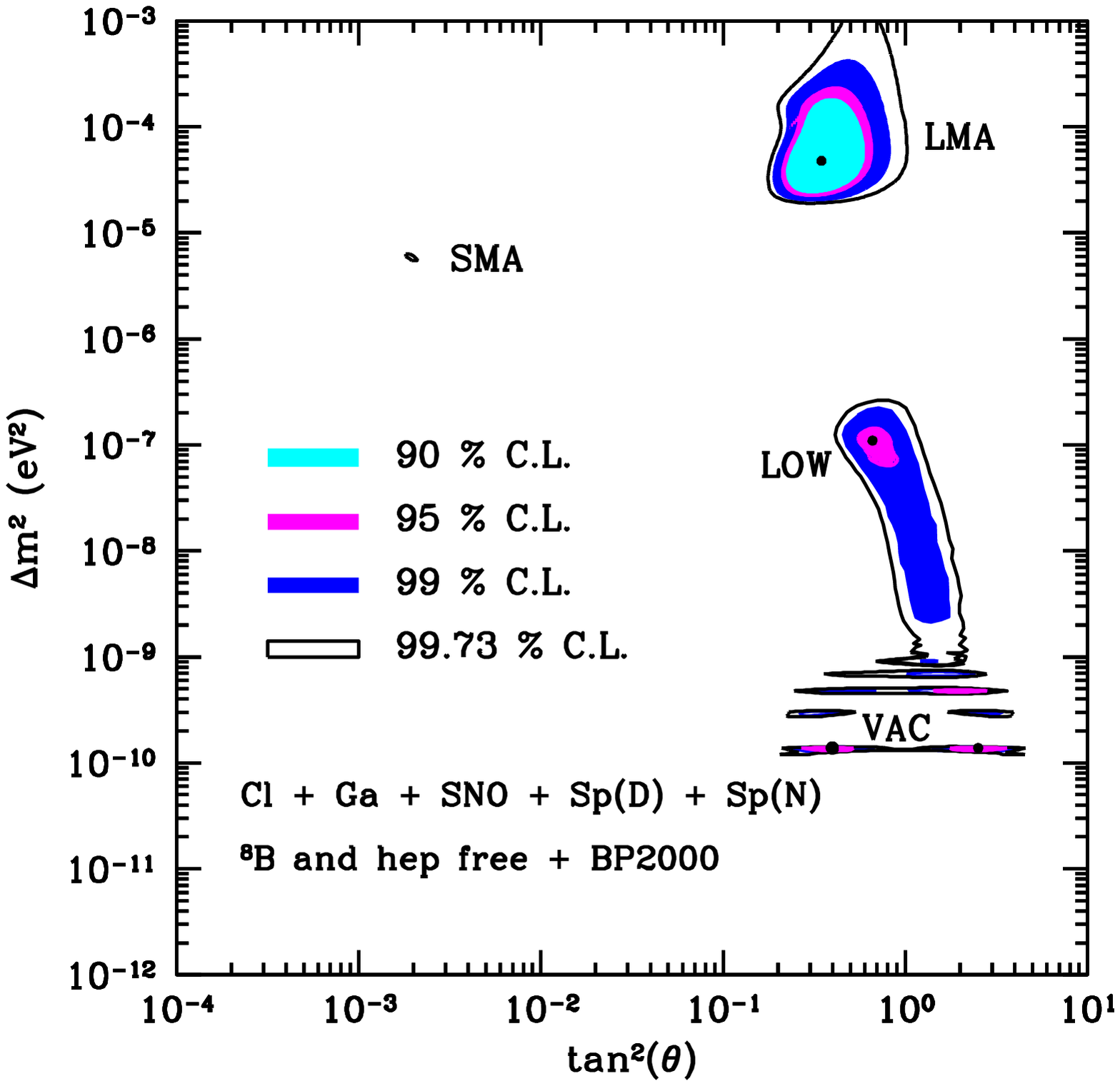}
\caption{
Global solutions from free flux analysis.   The  boron  and $hep-$
neutrino fluxes are considered as free parameters. 
The best fit points are marked by dark circles. The absolute
minimum of the $\chi^2$ is in the LMA region. 
The allowed regions   are shown at 90\%, 95\%, 99\% and 99.73\% C.L.  
with respect to the global minimum  in  the LMA region.
}
\label{global} 
\end{figure}

In Fig.~\ref{global} we show contours of constant (90, 95, 99, 99.73 \%) 
confidence level with respect of the
absolute minimum in the LMA region. Following the same procedure 
as in \cite{bks01}  the contours have been constructed in the following
way. For each point in the $\Delta m^2$, $\tan^2 \theta$ plane we find
minimal value $\chi^2_{min}(\Delta m^2, \tan^2 \theta)$ 
varying $f_B$ and $f_{hep}$. 
We define the contours of constant confidence level by the condition 
\be 
\chi^2_{min}(\Delta m^2, \tan^2 \theta) = \chi^2_{min} (LMA) + 
\Delta \chi^2~,
\label{delta}
\ee
where $\chi^2_{min} (LMA) = 29.3$ is the absolute minimum in the LMA
region and $\Delta \chi^2$ is taken for two degrees of freedom.

To clarify the role of the $hep-$neutrino flux we present in
Table \ref{Table2} the result of the $\chi^2$ analysis when 
$f_{hep} = 1$  and treat only $f_B$
as a free parameter.  As follows from the Table \ref{Table2}, 
fixing the flux
$f_{hep}$ leads to rather small change of the oscillation parameters. 
At the same time, the constraint $f_{hep} = 1$ lowers  a 
goodness of the fit of solutions which imply large value of $f_{hep}$
in the free flux analysis.  We get $\Delta \chi^2 = 1.8$ for VAC(sterile),
$\Delta \chi^2 = 1.4$ for SMA and $\Delta \chi^2 = 1.4$ for
VAC(active). Notice that now VAC(sterile) is shifted to the fourth 
position. 

According to Ref.~\ref{hepfl} the calculated $hep-$ neutrino  flux has 
about 20\% uncertainty. Therefore  solutions which require large
$f_{hep}$ (2 - 9)  are disfavored and the results of the fit with
$f_{hep}= 1$ look more relevant. \\

\begin{table}
\caption[]{Best-fit values of the parameters $\Delta m^2$, 
$\tan^2\theta$ and  $f_B$ 
from the global analysis with $f_{hep} =1$. 
The minimum $\chi^2$ and
the corresponding g.o.f. are given in the last two columns. 
The number of degrees of freedom is 39:
4 rates (Homestake,  SAGE, Gallex/GNO, SNO) + 38 SK spectral points  -
3 parameters.} 
\vskip 0.5cm
\begin{tabular}{l c c c c c}
\hline
Solution & $\Delta m^2$/$\rm eV^2$ &  $\tan^2\theta$ & $f_B$ &
$\chi^2_{min}$ & g.o.f. \\ 

\hline

LMA      & $5.0\times 10^{-5} $ &     0.36   &  1.1 &   30.1  & 0.85\\

VAC      & $1.4\times 10^{-10} $ & 0.363 (2.7) & 0.54 &  33.4 & 0.72 \\

LOW      & $1.1\times 10^{-7} $ &   0.69 &  0.86   & 34.5 & 0.67 \\
 
SMA      & $5.5\times 10^{-6} $ &  $1.9 \times 10^{-3}$ & 1.04 &  42.2 &
0.33\\

Just So$^2$ & $5.5\times 10^{-12}$ & 1.0 &  0.44 &  46.4 & 0.19 \\

VAC(sterile) & $1.4\times 10^{-10} $ & 0.35 (2.9) & 0.55 & 36.9 & 0.57 \\


SMA(sterile) & $3.8\times 10^{-6}$ &$4.2\times 10^{-4}$ & 0.52 & 
48.6 & 0.14 \\



\hline 

\end{tabular}
\label{Table2}
\end{table}   

\subsection{ SSM restricted global fit}

In order to check the significance of the SSM restriction on the boron
neutrino flux we have performed the fit to the data adding to the
$\chi^2$ sum in Eq.~(\ref{chi-def}) the term
\be
\left(\frac{f_B - 1}{\sigma_B}\right)^2~,
\label{boron-chi}
\ee
where $\sigma_B = 0.18$ is the average (of the upper and lower) 
$1\sigma$ theoretical error of the flux in BP2000 model \cite{ssm}.
With the term (\ref{boron-chi}) included in our global $\chi^2$ we, in
a way, treat the SSM prediction for the boron neutrino flux as independent
``measurement'' and consider it as an additional degree of
freedom.  This 
procedure makes sense   because a significant contribution to
the error in SSM determination of the boron neutrino 
flux comes from {\it the measurements} of the $p-Be$ cross-section.

One ``technical" remark is in order. Our approach differs from
analyses where the boron neutrino flux is taken as a theoretical
prediction from the SSM \cite{fogli,valencia}. In the latter case the
term (\ref{boron-chi}) is absent, the central SSM value is used in the
predictions of observables  and the theoretical errors on the boron
neutrino flux are added to the experimental errors.

\begin{figure}[ht]
\centering\leavevmode
\epsfxsize=1.0\hsize
\epsfbox{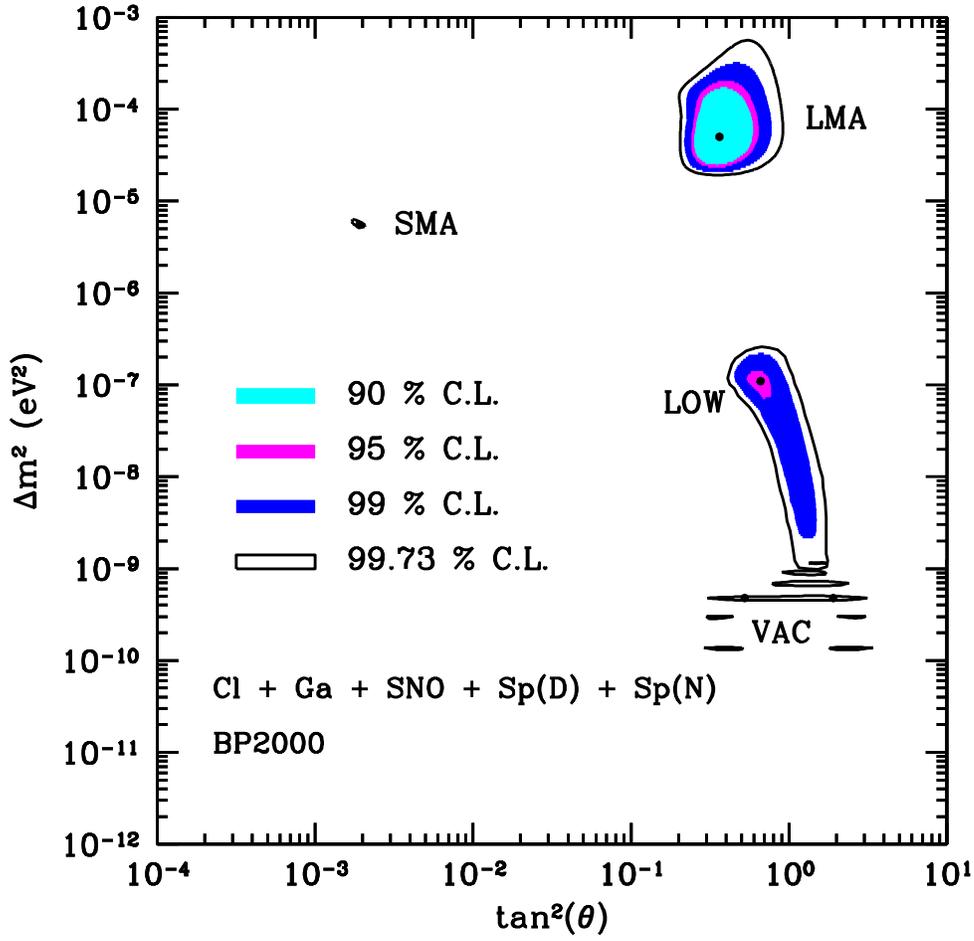}
\caption{
Global solutions for  the SSM restricted boron 
neutrino flux.   $hep-$ neutrino flux is free. 
The best fit points are marked by dark circles. The absolute
minimum of the $\chi^2$ is in the LMA region. 
The allowed regions   are shown at 90\%, 95 \%, 99\% and 99.73\% 
with respect to the global minimum  in  the LMA region.
}
\label{global-ssm} 
\end{figure}

In Table~\ref{Table3} and Fig.~\ref{global-ssm} we present the results of
the fit with SSM constrained boron neutrino flux. 
As expected, the most significant
changes in comparison with the free flux analysis (Table~\ref{Table1})
appear in those solutions and regions of the oscillation parameters
which imply strong deviation of $f_B$ from 1. Our results show that
mostly the best fit points, as well as the $\chi^2_{min}$ and goodness
of the VAC(active), Just-so$^2$, VAC(sterile) and SMA(sterile) solutions
are affected. Thus, for the VAC(active) solution 
$\Delta \chi^2_{min} = 3.4$, for VAC(sterile): 
$\Delta \chi^2_{min} = 6.2$. Moreover,  the best fit VAC solution shifts 
to another point of oscillation parameters in agreement with results of
other groups. 
There is no significant change of the LMA, LOW  and SMA
best fit points and goodness of the fit.

\begin{table}
\caption[]{Best-fit values of the parameters $\Delta m^2$, 
$\tan^2\theta$, $f_B$ and $f_{hep}$ 
from the SSM restricted global analysis. 
$f_{hep}$ is considered as a free parameter. 
The minimum $\chi^2$ and
the corresponding g.o.f. are given in the last two columns.
The number of degrees of freedom is 40:
4 rates (Homestake,  SAGE, Gallex/GNO, SNO) + 38 SK spectral points  + 
1 ($f_B$)  - 3 parameters.
}
\vskip 0.5cm

\begin{tabular}{l c c c c c c}

\hline 

Solution & $\Delta m^2$/$\rm eV^2$ &  $\tan^2\theta$ &
$f_B$ & $f_{hep}$ &  $\chi^2_{min}$ & g.o.f. \\

\hline 

LMA      & $5.0\times10^{-5}$ &     0.36   & 1.10  & 4.0 &   29.5 & 0.84\\

SMA      & $5.5\times10^{-6} $ &  $1.9 \times 10^{-3}$ &  1.10 & 4.0 &
        41.2 & 0.33\\

LOW      & $1.1 \times10^{-7} $ &   0.66  & 0.88 & 2.0   & 34.7 & 0.62 \\

VAC      & $4.8 \times 10^{-10} $ & 1.9 & 0.72 & 0.0 &  35.4 & 0.59 \\

Just So$^2$ & $5.5\times 10^{-12} $ & 1.4 & 0.44 & 1.0  &   55.4 & 0.03 \\

VAC(sterile) & $1.4\times 10^{-10} $ &2.63 & 0.55 & 8.0 & 41.3 & 0.33 \\

SMA(sterile) & $4.0\times 10^{-6}$ & $4.8\times10^{-4}$ & 0.54 & 2.0 &
54.5 & 0.04 \\

\hline 

\end{tabular} 
\label{Table3}
\end{table}

In Fig.~\ref{global-ssm} we show the contours 
of constant (90, 95, 99, 99.73 \%) confidence level 
for two degrees of freedom with respect
to the absolute minimum in the LMA region.\\

\subsection{Analysis of ``Rates Only" }

To clarify the relative significance of the total rates and the SK
spectrum in the global analysis of the data we have performed a fit to
the rates in the 4 experiments: Homestake, SAGE, GALLEX/GNO, SNO and
SK.  In this analysis we use the total rate of events in SK but  do
not use the SK energy spectrum of the recoil electrons.  The results of
the $\chi^2$ test are summarized in the Table 4.

Notice that with the SNO result the SMA solution does not give anymore
the best fit to the ``rates only'', in contrast with pre-SNO analyses.
The best fit is obtained in the VAC solution region.  However,
parameters of this solution differ significantly from the parameters
of the global solution when spectral data are included.

The ``rates only" fit shifts the LMA region to smaller 
$\Delta m^2$.  The LOW solution is practically unchanged.  Now
VAC(sterile) and LOW have low goodness of fit.

Notice that in the absence of the spectral data the 
value of $\chi^2_{min}$ is comparable or larger than 
the number of d.o.f. .

\begin{table}
\caption[]{Best-fit values of the parameters
$\Delta m^2$ and $\tan^2\theta$ from  the ``Rates Only'' analysis.  
The minimum $\chi^2$ and
the corresponding g.o.f. are given in the last two columns.
The number of degrees of freedom is 3:
5 rates (Homestake,  SAGE, Gallex/GNO, SNO, SK) - 2 parameters.}

\vskip 0.5cm

\begin{tabular}{l c c c c}
\hline
Solution & $\Delta m^2$/$\rm eV^2$ &    $\tan^2\theta$ & $\chi^2_{min}$ &
 g.o.f. \\ 

\hline
LMA      & $2.9\times 10^{-5}$  &     0.36   &     3.55  & 0.31\\

SMA      & $7.9\times 10^{-6}$  &  $1.4 \times 10^{-3}$ & 5.1 & 0.16\\

LOW      & $1.0\times 10^{-7}$  &   0.66    & 7.9& 0.05 \\

VAC      & $7.9\times 10^{-11}$ & 3.5 &   2.24 & 0.52 \\

Just-so$^2$  & $5.5\times 10^{-12}$  & 2.0  &16.4  & 0.0009  \\


SMA(sterile) &$4.4 \times 10^{-6}$  & $1.0 \times 10^{-3}$ & 17.4 &
0.0006\\


VAC(sterile) &$1.0 \times 10^{-10}$  & $0.35$ & 6.4 & 0.094\\
\hline

\end{tabular}
\end{table}  

\subsection{Pull-off diagrams}

In order to check the quality of the fits we have calculated
predictions for the available observables in the best fit points of
the global solutions found in the free flux analysis (see Table 5).
Using these predictions we have constructed the ``pull-off" diagrams
(fig.~\ref{pull}) which show deviations, $D_K$, of the predicted
values of observables $K$ from the central experimental values
expressed in the $1\sigma$ unit:
\be
D_K \equiv \frac{K_{bf} - K_{exp}}{\sigma_K}, ~~~~ 
K \equiv Q_{Ar},~Q_{Ge},~ R_{CC},~ R_{\nu e},~ A_{DN}^{SK}.   
\label{pull}
\ee 
Here $\sigma_K$ is the one sigma standard deviation for a given
observable $K$. We take the experimental errors only: $\sigma_K =
\sigma_K^{exp}$. The theoretical errors are related mainly to
the uncertainty in the boron neutrino flux. Since $f_B$ is treated as
a free parameter we do not take into account its theoretical errors.
The remaining theoretical errors are small and strongly correlated in
$Q_{Ar}$, $R_{CC}$ and $R_{\nu e}$.

\begin{table}
\caption[]{Values of the total rates in $Cl-$ , $Ga-$, SK and SNO 
experiments in the best fit points of global solutions found in the 
free flux analysis. The rates in the
radiochemical experiments are given in SNU. For SuperKamiokande and
SNO the ratios of the best fit rates  to the rates predicted in the SSM is
given.}

\vskip 0.5cm

\begin{tabular}{l  c c c c}

\hline
Solution  &  $Cl$ & $Ga$ & $SK$ & $SNO$ \\
\hline

LMA   & 2.89 & 71.3 & 0.452 & 0.323 \\

SMA   & 2.26 & 74.4 & 0.463 & 0.396 \\

LOW   & 3.12 & 68.5 & 0.446 & 0.368 \\

VAC   & 3.13 & 70.2 & 0.423 & 0.364 \\

Just So$^2$ & 3.00 & 70.8 & 0.434 & 0.434 \\\

SMA(sterile) & 2.93 & 75.5 & 0.435 & 0.445 \\
 
VAC(sterile) & 3.24 & 69.9 & 0.414 & 0.381 \\

Just So$^2$ & 3.01 & 70.9 & 0.434 & 0.435 \\

\hline
\end{tabular}
\end{table}

The diagrams are good diagnostics of the fit. They allow one to pin
down problems that some of the solutions have and to elaborate criteria
for further checks.

According to Fig.~\ref{pull} only the LMA  solution does not 
have strong deviations  of predictions from the experimental results. 
LOW, VAC and  SMA solutions give somewhat worser fit to the data. 
The fit from other solutions is very bad. 

The pull-off diagrams give some clarification to  a common worry, namely
that the high 
statistics SK experiment (in particular its spectral data) overwhelms
the rates data in the global analysis. In particular, solutions, which
are strongly disfavored by the rates, give a good fit when spectral
data are included.  

This problem can be approached in a different way: 
one can use just one parameter, {\it e.g.}, the first moment, which
describes possible distortions of the recoil electron energy spectrum
\cite{css}. \\

\section{Global solutions: properties and implications}

Here we evaluate the status of  global solutions using the
following criteria:

1. Goodness of the global fit in the free flux analysis.

2. Deviation of the boron and $hep-$neutrino  fluxes found in the free
flux analysis from the SSM values fluxes, 
that is,  the deviation of $f_B$ and $f_{hep}$ from 1.

3. Goodness of the fit in the SSM restricted analysis  and in the
``Rates only'' analysis.

4. Stability of the solution with respect to variations of the analysis.

5. Quality of the fit of the individual observables; features of the
pull-off diagram. A deviation by more than $3\sigma$ for some
observables is a clear signal for trouble.

We identify solutions which pass all of these criteria.

\subsection{The best fit solution: LMA}

SNO further favors the LMA solution \cite{lma99}.  In all global
analyses, in which the SK spectral data is included, LMA gives minimal
$\chi^2$. The best fit point from  the free flux analysis is:
\be
\Delta m^2 =  4.8 \cdot 10^{-5}~ {\rm eV}^2, ~~~\tan^2 \theta = 0.35, ~~~
f_B = 1.2,~~~ f_{hep} = 4.0~. 
\label{bfparam}
\ee 
Large $f_{hep}$ is needed to account for some excess of the SK events
in the high energy part of the spectrum.  For $f_{hep} = 1$, values of
the oscillation parameters are practically the same as in the free flux
analysis (see Table 2), and
the goodness of the fit is even slightly higher.   The boron
neutrino flux is 10\% higher than central value in the SSM: 
$F_B = f_B \cdot F_B^{SSM} = 5.66 \cdot 10^6$ cm$^{-2}$ c$^{-1}$ 
being however within
$1\sigma$ deviation.  This flux is rather close to the central value
extracted from the SNO and SK data \cite{sno}.

The fit of the data with the SSM restricted $f_B$ 
gives minimum of $\chi^2$ at practically the same values of 
parameters as in (\ref{bfparam}).

The values of the oscillation parameters (\ref{bfparam}) found here
are very close to the values found by other groups \cite{fogli,valencia}. 
This shows that 
the solution is robust and doesn't change with the type of analysis.

Notice that the SNO data lead to a shift of the best fit  point 
(as well as the whole region) to larger values 
of  $\tan^2 \theta$ as was discussed in the introduction.   

According to  the pull-off diagram, the LMA solution reproduces 
observables at $\sim 1\sigma$ or better. The largest deviation is for
the $Ar-$production rate: the solution predicts $1.1\sigma$ 
larger rate than the Homestake result. 

The ``Rates only" analysis shifts the best fit point to 
smaller $\Delta m^2$. \\ 

Considering  the allowed regions at different 
confidence levels we find the following: 

1).  $\Delta m^2$ is rather sharply restricted from below 
by the day-night asymmetry of the  SK  event rate: 
$\Delta m^2 > 2 \cdot 10^{-5}$  eV$^2$ at 99.73\%  {\rm C.L.} . 

2). The upper bound on $\Delta m^2$ is of great importance 
for future experiments, 
in particular for the neutrino factories \cite{nufac}. 
We find from the free flux fit (CHOOZ bound is not included)  
\be
\Delta m^2 \leq \left\{
\begin{array}{ll}
1.9 \times 10^{-4} ~ {\rm eV}^2,   &~~ 90\% ~ {\rm C.L.} \\
2.3 \times 10^{-4} ~ {\rm eV}^2,   & ~~95\% ~ {\rm C.L.}\\
4.3 \times 10^{-4} ~ {\rm eV}^2,   & ~~99\% ~ {\rm C.L.} \\
\end{array}
\right. . 
\label{up-on-dms}
\ee
Similar results can be obtained from the analysis 
in Ref.~\cite{fogli} where also CHOOZ data~\cite{CHOOZ} 
have been taken into account. 
The CHOOZ bound becomes important for larger than  $\Delta  m^2 \sim
8\times10^{-4} {\rm eV}^2$ where it modifies the $3 \sigma$ contour.

The fit with SSM restricted $f_B$ gives stronger bound on $\Delta
m^2$. Instead of the limits in Eq. (\ref{up-on-dms}) we get the upper
bounds $1.7 \cdot 10^{-4}$, $2.1 \cdot 10^{-4}$, and $3.1 \cdot
10^{-4}$ eV$^2$ for 90, 95 and 99 \% C.L. correspondingly.

3). The SNO and  SK results  (evidence of $\nu_{\mu}$, $\nu_{\tau}$
appearance) give an important {\it lower} limit on  mixing: 
\be
\tan^2 \theta > 0.2~~~~   99\% ~{\rm  C.L.}. 
\label{theta}
\ee

4). Maximal mixing is allowed only at the $\sim 3\sigma$ level:
\be
\tan^2 \theta \geq 1~~~~ {\rm for} ~~ \Delta m^2   
= (4 - 10) \cdot 10^{-5}~~ {\rm eV}^2,  ~~~~~~~  99.73 \%~~ {\rm C.L.} .  
\label{maxmix}
\ee   
In spite of the shift of the best fit point to larger values of
$\Delta {m}^2$ the C.L. for acceptance of maximal mixing is not
lower than it was before the SNO result.  The reason is that the SNO
rate corresponds to a survival probability smaller than 1/2, which
disfavors maximal mixing.

Similar result follows from the analysis in Ref.\cite{fogli} where it
was found that maximal mixing (at $3 \sigma$ level) is allowed for
$\Delta {m}^2 = (4 - 20) \cdot 10^{-5}$ eV$^2$.\\  

In the fit with the SSM restricted $f_B$, maximal mixing is even more
disfavored: $\tan^2 \theta < 0.9 $ at 99.73 \% C.L..

\subsection{Large or Small? The fate of SMA}

The fate of the SMA solution, the only solution which is based on
small mixing, is of great importance for future developments in both 
theory and experiment.

We find that the SMA solution does not appear at $3 \sigma$ level 
(with respect to the global minimum) in the free flux fit. The best fit
point parameters are: 
\be
\Delta m^2 =  6.0 \cdot 10^{-6}~ {\rm eV}^2, ~~~\tan^2 \theta = 0.0019,
~~~
f_B = 1.12,~~~ f_{hep} = 4.
\label{bfparamSMA}
\ee 
SNO shifts the local minimum to substantially larger 
mixing angles in comparison with  the pre-SNO result. This is a 
consequence of  appearance of the 
$\nu_{\mu} / \nu_{\tau}-$ flux, which implies large transition probability 
and therefore large mixing angles.  
At such a large $\tan^2 \theta$  one expects significant distortion of the 
boron neutrino spectrum (see sect. 4). 

Important feature of the solution is large flux of the $hep-$neutrinos. 
It is this large flux which,  together with correlated 
systematic error and $f_B > 1$, makes possible to get a reasonable
description of the SK 
energy spectrum. 
If the SSM value is taken for the $hep-$neutrino flux (see Table 2)  
the $\chi^2$ increases by $\Delta \chi^2 = 1.4$.

These results are in agreement with those obtained in \cite{fogli}.
In \cite{valencia},  the SMA
is accepted at lower than 99\% level. Surprizingly, in this analysis 
the SNO result shifts mixing to even smaller values:  
the best fit value from
\cite{valencia} is at  $\tan^2 \theta = 4\cdot 10^{-4}$.  


Before the SNO result, the SMA solution was always giving the best fit
to the total rates. Inclusion of the $CC-$event rate measured by SNO
moves the SMA solution to third position, after VAC and LMA.
Furthermore, the fit is no longer good: $\chi^2_{min} = 5.1$ for 3
d.o.f..  From the pull-off diagram we find that there is a tension
between the SNO and Homestake rates: The SNO data ($CC-$rate) requires
rather small survival probability for boron electron neutrinos.
Furthermore Gallium experiments imply  strong suppression of the
Beryllium neutrino flux.  
This leads to low ($1.9 \sigma$) $Ar-$ production
rate. At the same time, according to fig.~\ref{pull} the $CC-$event
rate is $1.7 \sigma$ higher than the SNO result.
 
For mixing angle corresponding to the best fit point
(\ref{bfparamSMA}) one expects significant regeneration effect in the
core-night bin \cite{core-bin} due to parametric enhancement of
oscillations for the core crossing trajectories
\cite{par-res}. However, the day and the night spectra we used in our
analysis are not sensitive to this feature.  The zenith angle
distribution of events measured by SK does not show any core
enhancement \cite{SK}. Inclusion of this information in the
global analysis will further disfavor the SMA solution.\\

\subsection{LOW: next best?}

For the best fit point of the free flux analysis we get:
\be
\Delta m^2 =  1.1 \cdot 10^{-7}~ {\rm eV}^2, ~~~\tan^2 \theta = 0.66, ~~~
f_B = 0.88,~~~ f_{hep} = 2.0. 
\label{bfparamLOW}
\ee
If the $hep-$neutrino flux is fixed at its SSM value, $f_{hep} = 1$,
the best fit point shifts to larger  mixing: 
$\tan^2 \theta  =  0.69$. 
Notice that
the solution implies $\sim 1 \sigma$ lower boron neutrino flux than in
SSM.

In the analysis with SSM restricted $f_B$ the best fit point is the same
as in  (\ref{bfparamLOW})

The LOW solution gives rather poor fit of the total rates $\chi^2
=7.9$ for 3 d.o.f. (see Table 4).  In the best fit point we get $2 
\sigma$ larger $Ar-$production rate and  $1.9 \sigma$ lower
$Ge-$production rate. It is this
case when the overwhelming spectral information ``hides" 
some problems with
total rates in the global fit.  The use of a single parameter for
description of the spectrum distortion gives much lower goodness of
the fit for the LOW solution as compared with the LMA solution
\cite{css}.\\

\subsection{VAC - oscillation solution is back ?}

As anticipated from the  comparison of the pre-SNO
predictions 
with the SNO result the VAC solution  improves  its status. 
Indeed, in the best fit point of the free flux analysis:  
\be
\Delta m^2 =  1.4 \cdot 10^{-10} {\rm eV}^2, 
~~~\tan^2 \theta = 0.40~(2.5), ~~~
f_B = 0.53,~~~ f_{hep} = 6.0.
\label{vac1}
\ee  
$\chi^2$ is even lower than the LOW solution has. 
The solution with parameters (\ref{vac1}) has been found in the
pre-SNO analysis \cite{bks01}, but before the SNO result the goodness
of the fit was substantially lower in comparison with other
solutions.

The solution gives very good description of the SK energy  spectrum.  
The $\chi^2$ is substantially smaller than the number of degrees of
freedom.   The solution reproduces
rather precisely the bump in both the night and the day spectra at (7
- 8) MeV and the dip at (11 - 12) MeV. (This features can be well seen
in fig. 4e  from the Ref.~\cite{bls}.)  
The bump in the spectrum originates from the first maximum of the
oscillation probability which corresponds to the  oscillation phase  
$2\pi$. Above the bump the probability decreases with energy and 
the increase of $R_{\nu e}$ at 
$E > 12$ MeV is due to large flux of the $hep-$neutrinos.

Notice that an excellent description of the spectrum requires 
non-maximal mixing, otherwise the distortion is very strong. 
Value of $\sin^2 2\theta$ which  
immediately determines the depth of oscillations should be  
about $0.8$.  Then  to compensate for 
rather large survival probability and to explain the SNO result one
needs to assume a small ($f_B \sim 0.5$) original boron 
neutrino flux.

However, there are several problems with this solution.  It requires
substantially ($ \sim 3 \sigma$) lower original boron neutrino flux 
than in SSM and substantially higher original $hep-$neutrino flux:
$f_{hep} = 6$.  
In the fit with $f_{hep} = 1$ the $\chi^2$ increases 
by $\Delta \chi^2 = 1.4$. 

For this solution one predicts the seasonal asymmetry due to
oscillations $A \approx - 0.6 A_0$, where $A_0 \sim 7 \%$ is the asymmetry 
due to the geometrical factor only ($1/R^2$ change of the flux). 
Thus one expects a suppressed seasonal asymmetry in contrast with
observations.

The solution gives rather good fit to the  rates. 
Although  in the ``Rates
only" analysis the best fit point shifts to a different island in
parameter space.  According to the pull-off diagram, the solution
predicts $2.1 \sigma$ higher $Ar-$production rate  
and $2.6 \sigma$ lower SK rate, and no day-night
asymmetry. 

The SSM restricted global fit shifts the best fit point to an 
``island" centered around
\be
\Delta m^2  =  4.8  \cdot 10^{-10}~ {\rm eV}^2, ~~~~\tan^2 \theta = 1.9.  
\label{vac2} 
\ee
Now $f_B = 0.72$. This solution was found by other groups too.  
The  solution in the same ``island" of the oscillation parameter space  
already appeared before (after 508 days of SK operation)
when a significant excess of events at the high energy part of the
spectrum was observed.  The solution was later excluded by SK data on
the spectrum. After the SNO result it re-appeared again.
The solution can reproduce some bump at $E \sim 8$ MeV and 
increase of $R_{\nu e}$ at the high energies. 

Thus,  in the VAC region there are three local minima 
(two of them are degenerate) with rather close
$\chi^2$. Small variations in the analysis shift the best fit VAC
point from one minimum to another.   

Although the VAC solution provides a very good fit to the data in the
free flux global analysis, it does not pass additional criteria of
quality. It requires very strong deviations of the boron and hep
neutrino fluxes from the SSM values.  The goodness of the fit becomes
substantially worse when the SSM restrictions are imposed on these
fluxes. There are significant deviations in the pull-off diagram.\\

The Just-so$^2$ solution looks extremely unlikely in the light of the SNO
result. It gives a very poor fit to the rates. In fact, the fit is so bad
that even the flat spectrum that it predicts, in agreement with the SK
measurement, is not sufficient to make it plausible. Given the excellent
fit of the LMA solution, the Just-so$^2$ solution is ruled out
at 3$\sigma$ C.L.. This result is in agreement with Ref.\cite{fogli}.  
Note that if SNO fails to
observe a large day-night asymmetry, the situation might change and the
Just-so$^2$ solution might reappear at about $\sim 3 \sigma$
C.L..

\subsection{How large is large mixing?}

This question is crucial for theory.  In a number of approaches to the
bi-large mixing one gets mixing of the electron neutrino which is very
close to maximal mixing (see
\cite{max} for a general discussion). 

In the LMA region we find from the free flux analysis: 
\be
\tan^2 \theta < \left\{
\begin{array}{ll}
  0.68~~  &  95 \% ~~    {\rm C.L.} \\ 
  0.82~~  &  99 \% ~~     {\rm C.L.} \\ 
  1.05~~  &  99.73 \%  ~~   {\rm  C.L.}\\
\end{array}
\right.
\label{theta-up}
\ee 
and,  as we discussed in sect. 3.1,  maximal mixing is allowed  
at $\sim 3 \sigma$ level for $\Delta m^2 = (4 - 10) \cdot 10^{-5}$ eV$^2$.  
In the SSM restricted analysis the bounds become stronger.  

In the LOW region we find $\tan^2 \theta < 0.9$ at 95\% C.L.. 
Maximal mixing is accepted at slightly lower than 99\% C.L. . 
in the range $(6 \cdot 10^{-9}    - 3 \cdot 10 ^{-7})$ eV$^2$ 
which covers LOW and  the so called QVO (Quasi-Vacuum  Oscillation) 
regions.

Maximal mixing is the best fit value of the Just-so$^2$ solution.   
Although  this solution is ruled out at $3 \sigma$ level in the global 
analysis. 
Notice, the fit of the data in the VAC region implies 
significant deviation from maximal mixing.\\   

\subsection{Do pure sterile solutions exist?}

The best (and the only accepted at $3 \sigma$ level) pure sterile
solutions is the VAC(sterile), 
as could be expected from our pre-SNO analysis \cite{bks01}. 
In the free  flux analysis the solution appears at  90\%  C.L. 
already with the best fit point: 
\be
\Delta m^2 =  1.4 \cdot 10^{-10}~ {\rm eV}^2, 
~~~\tan^2 \theta = 0.38~ (2.6), ~~~
f_B = 0.54,~~~ f_{hep} = 9.
\label{bfparamVAC}
\ee
Partially,  the difference between the SK and SNO rates is explained
by distortion of the spectrum: the suppression increases
with energy and therefore the higher threshold at SNO leads to lower
averaged survival probability.  However, mainly the fit ``shares" 
the deviations between SNO and SK: the solution requires $1.2 \sigma$
higher  SNO rate and 
significantly lower boron neutrino flux which gives a $\nu e-$
scattering rate $3.0 \sigma$ below the SK result (see
fig.~\ref{pull}).  
The free flux analysis without SK total rate does
not locate this problem immediately.  
In principle,
it should reappear  in the fit of the spectrum with 
similar  $\Delta \chi^2$ since the solution fits the shape of the
spectrum rather well.

Notice that the parameters of this solution  are very close 
to the parameters of VAC(active) apart from the two times 
larger $hep-$flux. So this solution leads to the same type of the spectrum
distortion as the one described in sect. 3.4  with the bump at 
$E = 7 - 8$  MeV and the dip at $E = 11 - 12$  MeV.

In addition to a strong deviation of the $\nu e$ scattering event rate
from the SK result, the solution predicts $2.1\sigma$ larger
$Ar-$production rate.  The fit worsens when SSM restrictions 
are imposed.  For the SSM value of the $hep-$neutrino flux 
the best fit point shifts to a smaller mixing angle
and  the $\chi^2$ increases by $\Delta \chi^2 = 1.8$. 
In the analysis with SSM restricted $f_B$, the
goodness of the fit drops further: $\Delta \chi^2 = 6.2$. 
In the fit of the rates the solution shifts to smaller 
$\Delta m^2$ where a description of the SK spectra becomes bad. 
Thus,  the solution  does not pass additional quality tests.

The SMA(sterile) solution gives a very bad fit since it leads (in contrast
with the VAC solution) to a distortion of the boron neutrino spectrum 
with suppression which  weakens with increase of energy.  
This solution contradicts the SNO result: 
$3.3\sigma$ higher rate is predicted.\\

Concerning sterile solutions one remark is in order. 
Better fit of the data can be obtained if more than 1 sterile neutrino
participate in the conversion. Such a  possibility is realized 
when  solar neutrinos convert to the so called  
``bulk" neutrinos  which propagate both in usual  and  in (large) 
extra space dimensions ~\cite{dvali}.  
From the four dimensions point of view 
the solar $\nu_e$ is transformed to several Kaluza-Klein mode of the
bulk neutrino  which show up as sterile neutrinos. 
In this case one can reproduce 
(with low $\chi^2$)   
the SK  spectral data, and at the same time have better
descriptions of the total rates (see~\cite{bulk}  for recent analysis).

\section{SNO: Predictions for the next step}

Forthcoming results from SNO will include measurements of the
day-night asymmetry and a more precise determination of the electron
energy spectrum (higher statistics and lower energy
threshold). Later results on the NC/CC ratio will be available.
Predictions for these observables have extensively been discussed
before \cite{B-K,M-P,flmp,zenith,bl-mom,bkl}. Here we sharpen the
predictions using the latest solar neutrino data. 
We also calculate the expected values  of the total event rate 
in the BOREXINO experiment which start to operate soon. 
\\

\subsection{Day-Night asymmetry}

We calculate  the D-N  asymmetry at SNO,  $A_{DN}^{SNO}$, defined as
\be
A_{DN}^{SNO} \equiv 2\frac{N - D}{N + D}
\label{dndef}
\ee  
for the events above the threshold $E^{th} = 6.75$ MeV.

We  compare the asymmetry of  the $CC -$ events at SNO with 
the asymmetry of the $\nu e-$events measured  at SK: 
$A_{DN}^{SK}$.  There are tree factors which can  lead 
to  substantially different SNO and SK asymmetries: 

A ``dumping'' factor, $\eta_{dump}$, describes the suppression of the
D-N asymmetry in the $\nu e-$ event rate due to the contribution of
the $\nu_{\mu}, \nu_{\tau}$ scattering to the signal.  We get
\cite{bks-10}
\be
A_{DN}^{SNO} \propto  \eta_{dump} A_{DN}^{SK}, 
\ee
where 
\be
\eta_{dump} = 1 + \frac{r}{(1-r) \bar P}~. 
\label{eta}
\ee
Here $r \equiv \sigma (\nu_{\mu})/ \sigma (\nu_{e}) $ is the ratio 
of cross-sections of the $\nu_{\mu}  e- $  and $\nu_{e} e - $ scattering, 
and $\bar P$ is the averaged survival probability. 
Notice that with decrease of $\bar P$ the damping factor increases. 

A second factor, $\eta_{thr}$, describes the effect of difference of
the energy thresholds: 5 MeV for SK and 6.75 MeV for SNO. It also
accounts for the larger minimum difference 
(the binding energy of the deutron, 1.44 MeV) 
between neutrino energy and electron
energy in SNO.

A third factor, $\eta_{reg}$, is related to the difference of the
geographical latitudes.  Since the Sudbury mine is at higher latitude, the
regeneration effect there is slightly weaker than at Kamioka.

The total difference between the SNO and SK asymmetries is the product
of these three factors.\\

Let us consider now the predictions for the asymmetries for individual
solutions.

{\it 1).  LMA- solution}. In Fig.~\ref{dn-l} we show dependence of
the asymmetry on $\Delta m^2$ for different values of $\tan^2 \theta$
from the allowed region.  The asymmetry decreases with increase of
$\Delta m^2$ as $ \sim 1/\Delta m^2$ for $\Delta m^2 > 4 \cdot
10^{-5}$ eV$^2$ and at larger $\Delta m^2$ the decrease is faster due
to effect of the adiabatic edge (see \cite{bks-corr} for details).
The dependence of asymmetry on $\tan^2 \theta$ is rather weak.  In the
best fit point we get
\be
A_{DN}^{SNO} = 7.2 \% ~~~~(E^{th} = 6.75 ~{\rm MeV}).    
\label{bfasymm}
\ee 
In the allowed region the asymmetry can take any value from
practically zero to 15\% at 90 \% C.L.. At $3 \sigma$ level it can
reach 20 \%.  The asymmetry is maximal at the smallest possible values
of $\Delta m^2$ and $\tan^2 \theta$ within the allowed region.  It
increases with the energy threshold due to an increase of the
regeneration factor: $f_{reg} \propto E/\Delta m^2$.

The expected asymmetry at SNO is substantially larger than at SK. In
the best fit point we get $A_{DN}^{SK} = 3.6 \%$, thus $A_{DN}^{SNO}
\approx 2 \cdot A_{DN}^{SK}$. This difference can be easily understood
considering the factors mentioned above. Indeed, for the best fit
point the survival probability can be estimated as $\bar P \approx
R_{SNO}/f_B$. Inserting numbers from the Table 5 we get from
Eq. (\ref{eta}) $\eta_{dump} = 1.6$.  In the LMA region the asymmetry
increases with $E_{th}$.  The threshold factor equals $\eta_{thr} \sim
1.2$, and $\eta_{reg}$ is close to 1.  As a result, the overall
enhancement factor at SNO is about 2.

{\it 2). LOW solution}. The dependence of the asymmetry 
on $\Delta m^2$ for different values of $\tan^2 \theta$ from the allowed
region is shown in Fig.~\ref{dn-low}. 
In the best fit point (of the free flux analysis)
\be
A_{DN}^{SNO} = 2.4 \% ~~~~(E^{th} = 6.75 ~{\rm MeV}).
\label{bfasymm}
\ee 
The asymmetry increases linearly with $\Delta m^2$ for $\Delta m^2 >
10^{-7}$ eV$^2$. It can reach 10 - 12 \% at the upper border of the
allowed ($3 \sigma$) region.  For $\Delta m^2 < 10^{-7}$ eV$^2$ the
asymmetry decreases faster with $\Delta m^2$ due to effect of the
non-adiabatic edge (see \cite{bks-corr} for details).  
It depends weakly  on the  mixing angle.

Let us emphasize that the D-N asymmetry in the best fit point of the
LOW solution is substantially smaller than the one in the LMA
region. Thus, an observation of asymmetry $A_{DN}^{SNO} > 5 \%$ will
strongly favor the LMA solution.

In the LOW region the D-N asymmetry at SNO is also larger than in the
SK detector.  However the difference here is not as  large as in the LMA
case.  In the best fit point we get $A_{DN}^{SK} = 2 \%$, which
results in $A_{DN}^{SNO} = 1.2 A_{DN}^{SK}$.  There are two reasons
for such a difference:

(i) The average survival probability is larger now (mixing angle is
larger): $\bar P \approx R_{SNO}/f_B = 0.42$ (see Table 5).  As a
consequence, the damping factor is smaller: $\eta_{damp} = 1.45$.

(ii) The difference in the thresholds works in the opposite direction
in comparison with the LMA case, thus suppressing the SNO
asymmetry. Indeed, in the LOW region the regeneration factor decreases
with {\bf $E/\Delta m^2$} and therefore the asymmetry decreases with
increasing energy threshold.  For $\eta_{thr} \sim 0.85$ we get a
total enhancement factor in agreement with the exact calculation.\\

{\it 3). SMA solution}. The dependence of the asymmetry on $\tan^2 \theta$ 
for different values of $\Delta m^2$ is shown in Fig.~\ref{dn-sma}. 
In the best fit point we get $A_{DN}^{SNO} = 2.6 \%$ and 
most of this asymmetry is collected from  the ``core" bin. 
The asymmetry increases fast with $\tan^2 \theta$. In the 
$3 \sigma$ allowed region it can be as large as  8\%.  

The SNO asymmetry is only slightly higher than the SK- asymmetry:
$A_{DN}^{SK} = 2.4\%$. In the SMA range due to strong dependence of
the survival probability on energy the relation between the SNO and SK
asymmetries is more complicated than in Eq. (\ref{eta}).

\subsection{Spectrum distortion}

As follows from previous studies \cite{bl-mom,bkl,bls}, in general the
sensitivity of the SNO measurements to the energy spectrum distortion
is higher than the sensitivity of SuperKamiokande due to better
correlation of the neutrino and the (produced) electron
energies. However, the present statistics in SNO is much lower than in 
SK, and the statistical errors are more than 2 times larger (in the
range 8 - 10 MeV we get 10 -12 \% at SNO as compared with 4 - 5 \% at
SK).
  
In our analysis we do not use the spectral information from
SNO. Instead, we present here a qualitative discussion comparing the
present SNO data with predicted spectra from different solutions. This
allows us to evaluate significance of the present and forthcoming SNO
results.

In Fig.~\ref{spectra} we show the expected spectra of events at SNO for
several
values of the oscillation parameters in the LMA region. In the high
energy part of the spectrum the distortion is due to the earth
regeneration effect as well as  contribution of the $hep-$ neutrino
flux. However, for $\Delta m^2 \leq (4 - 5)\cdot 10^{-5}$ eV$^2$ 
the regeneration effect is small and the turn up at $E > 11$ MeV is
due to increased ($f_{hep} = 5$) flux of the $hep-$neutrinos.  The
turn up of the spectrum at low energies is the effect of the adiabatic
edge of the suppression pit.  Recall that the ratio of the event rate
with oscillations to the event rate with no oscillations is determined
by the product of the survival probability times the factor $f_B$:
$R_{CC} \propto f_{B} (\Delta m^2) \bar{P} (\Delta m^2)$. With
increase of $\Delta m^2$ spectrum shifts to the adiabatic edge; for
$\Delta m^2 < 10^{-5}$ eV$^2$ the probability can be written as 
$$
\bar{P} \approx (\sin^2 \theta + f_{reg} + A), 
$$
where $A$ is the correction due to adiabatic edge and it is
proportional to the first moment $\delta T$: 
\be
A \propto \delta T \propto \cos 2\theta \left( \frac{\Delta m^2}{E}
\right)^2
\ee
(see Appendix of \cite{bks-corr} for more details).  The factor $\cos
2 \theta$ accounts for the disappearance of the distortion when the
mixing approaches maximal value.  With increasing $\Delta m^2$ the
size of the turn up and the distortion of the spectrum (first moment)
increase as $(\Delta m^2)^2$.  Notice that at the same time $f_B$
decreases to compensate for a total increase of the
survival probability.  The
distortion reaches a maximum at $\Delta m^2 \sim 1.5 \cdot 10^{-4}$ 
eV$^2$ when the
boron neutrino spectrum is in the middle of the adiabatic edge.  With
further increase of $\Delta m^2$,  the spectrum shifts out of the
suppression pit and the distortion decreases.  For $\Delta m^2 > 3
\cdot 10^{-4}$ eV$^2$ the spectrum is in the range where the
probability is determined basically by the averaged vacuum
oscillations: $P = 1 - \sin^2 2\theta /2$ and does not depend on 
the neutrino energy. Notice that the distortion of the spectrum 
weakens due to
integration over neutrino energy and folding with the energy
resolution function.  Moreover, the SNO sensitivity to the distortion
at low energies is weakened by the fast decrease of the cross-section
with  energy.

The curves shown in Fig.~\ref{spectra}a  illustrate this behavior of the
spectrum. The short dashed line corresponds to near maximal distortion. As
follows
from the figure it will be very difficult to establish the distortion
expected from  the LMA solution with SNO.  One needs to measure 
the spectrum down to $\sim 5$  
MeV and to increase substantially the statistics.  Thus, the prediction
for SNO is that no significant distortion should be seen in
forthcoming measurements.

In Fig.~\ref{spectra}b we show the expected spectra of events at SNO in
the best
fit points of the LOW, SMA and VAC(active) solutions.  
The  correlated systematic errors (mainly due to the error
in the absolute energy scale calibration) are not shown here. These
errors can affect the conclusions from the fit to the spectrum,
making the spectrum appear flatter.

In the best fit point of the LOW solution the Earth  
regeneration effect on
the shape of the spectrum is small.  The weak positive slope (positive
shift of the first moment) is due to the effect of the adiabaticity
violation (non-adiabatic edge of the suppression pit). The slope
increases with decreasing $\tan^2 \theta$ as well as $\Delta m^2$ (for
some analytical studies see \cite{bks-corr}). It will be difficult to
establish this distortion with SNO.

For the VAC (active) solution with low $\Delta m^2$,
Eq.~(\ref{vac1}), one predicts maximum of the ratio $R_{CC}$ at $E
\sim 6$ MeV. It corresponds to the first oscillation maximum of the
survival probability. The ratio decreases with increase of energy and
at $E > 11$ MeV the distortion flattens due to contribution of the
$hep-$neutrino flux.  Thus,  negative shift of the first moment (slope)
is expected. 
As follows from the figure, the distortion is at the border of
sensitivity of already present SNO data (correlated systematic errors
can partly improve an agreement between the data and the predictions).
Further increase of statistics and especially measurements of the
spectrum at lower energies will be crucial for discrimination of the
solution. Notice that the oscillation parameters are well fixed by the
SK spectrum, and no freedom exists, e.g., to change the distortion by
varying $\Delta m^2$. In particular, the position of the
maximum at $E \sim 6$ MeV is immediately determined by the maximum in
the SK spectrum at 8 MeV. The distortion can be identified by
comparison of the averaged ratio $R_{CC}$ at low and high energies:
$R_{CC}( < 9~ {\rm MeV})$  and $R_{CC}( > 9 ~{\rm MeV})$.

Similar distortion is expected for the VAC(sterile) solution, see
Fig.7c. For VAC (active) with high $\Delta m^2$ 
(see Eq.~(\ref{vac2})) one
expects flat distribution at low energies due to strong averaging
effect and a bump at high energies $E > 12$ MeV.

For the SMA (active) solution there is a strong positive shift of the
first moment (slope).  Correlated systematic errors improve the 
agreement with the experimental data. The present SNO data 
can already 
have some impact on this solution. The allowed region will drift to
smaller mixing angles where the distortion is weaker. The best fit
boron flux should decrease in order to compensate for the increase of
the survival probability.

For SMA(sterile) the distortion is very weak, as a consequence of very
small mixing angle. However for this solution the predicted spectrum
is systematically above the experimental points which corresponds to
$3.3 \sigma$ higher total rate.

In conclusion, the present SNO spectral data further favor the LMA and
LOW solutions. However, it will be difficult to establish with SNO the
weak spectral distortions expected for these solutions.  At the same
time the SNO measurements can be sensitive to strong distortion of the
spectrum predicted by already disfavored solutions such as VAC and
SMA.

\subsection{NC/CC ratio}

The reduced  neutral current 
event rate [NC] is defined as the ratio of the rates with and without
oscillations:  $[{\rm NC}] \equiv N_{NC}/N_{NC}^{SSM}$. 
Similarly, the reduced rate of the charged current event rate 
equals $[{\rm CC}] \equiv N_{CC}/N_{CC}^{SSM}$. 
We have calculated the ratio of the NC and CC reduced rates,  
[NC]/[CC], for global solutions found from  the free flux fit.  

For the active neutrino conversion the ratio equals 
\be
\frac{\rm NC}{\rm CC} = \frac{1}{\bar{P}} \approx \frac{f_B}{R_{SNO}}, 
\label{nc-ccratio}
\ee
where $\bar{P}$ is the effective (averaged) survival probability 
of the electron neutrinos. In the best fit points, using results 
for $f_B$ and $R_{SNO}$  from the Tables 1 and 5,  we find 
[NC]/[CC] $= 3.5$ (LMA), 2.4 (LOW), 2.8 (SMA), 1.5 (VAC), $\approx 1$
(Just-so$^2$) in agreement with results of numerical calculations 
(see fig.~\ref{nc-cc}).  

For the sterile solutions we have: 
\be
\frac{\rm [NC]}{\rm [CC]} = \frac{\bar{P}_{NC}}{\bar{P}_{CC}},  
\label{nc-cc-s}
\ee   
where $\bar{P}_{NC}$ and  $\bar{P}_{CC}$ are the effective survival
probabilities for the NC and CC samples correspondingly. For the 
NC sample the energy threshold is about 2.2 MeV, so that  ${P}_{NC}$ 
is averaged over  larger interval than ${P}_{CC}$. 
As a consequence, the probabilities  $\bar{P}_{CC}$ and   $\bar{P}_{NC}$
are in general different. 

Notice that the largest value of the ratio  
is expected for the LMA solution:  
\be
\frac{\rm [NC]}{\rm [CC]} = 3.5^{+2.4}_{-1.7} ~~~({\rm LMA})~. 
\label{nc-cc-lma}
\ee  
Then rather close predictions 
follow from the SMA and LOW solutions. 
Smaller value is predicted for VAC. 
For sterile solutions the ratio is close to 1. 

According to fig.~\ref{nc-cc}, there is a significant overlap of the 
$3\sigma$ regions of predictions. 
However, measurements of the ratio with better than 20 \% accuracy 
can significantly contribute to discrimination of the solutions. 

\subsection{BOREXINO rate}

For  global solutions  
we  have calculated the reduced event rate in the  BOREXINO 
experiment \cite{bor},  
$R_{borexino} \equiv N_{osc}/N_{SSM}$,   
where $N_{osc}$ and $N_{SSM}$ are the 
expected rates with  and without neutrino conversion. 
We averaged the effect 
over the year.

The rate can be estimated as 
$R_{borexino} = \bar{P}_{Be}(1 - r) + r$, where 
$\bar{P}_{Be}$ is the averaged 
survival probability for the $^7$Be- neutrino line. The probability  
$\bar{P}_{Be}$   has been calculated for the 
oscillation parameters from the $3\sigma$  allowed regions  
of solutions found from the free flux analysis.  
$r$ is the ratio of  cross-sections of the 
$\nu_{\mu}  e- $ and $\nu_e e-$ scattering at the   
the beryllium line.  
We have used the neutrino-electron
scattering cross-sections from Ref.~\cite{BKSirlin}. 

The results of calculations for six global solutions are shown in the
Fig.~\ref{borex}.  The rate is suppressed for  
the  solutions.  In particular, for the LMA solution we get 
$$
R_{borexino} = 0.66^{+0.11}_{-0.08}~~~~({\rm LMA}). 
$$
Similar suppression is expected for the 
LOW. The VAC solutions predict stronger suppression, the lowest  
rate is  expected for SMA. 
Notice that the range of predictions for SMA is very small 
due to smallness of the allowed region itself. 

There is a significant overlap  of the predicted  
intervars for LMA, LOW and VAC. 
Therefore,  it will be difficult to discriminate among these solutions 
using  just total rate measurements. 
However,  Borexino  has other
capabilities, e.g. it can detect strong  day-night effect in the  
case of the  LOW solutions \cite{bor-dn},  whereas strong seasonal 
variations are expected if  QVO or VAC solution is realized 
\cite{bor-seas}.

\section{Conclusions}

We have performed a global analysis of the solar neutrino data
including the charged current event rate measured by SNO.  We tested
the robustness of the global solutions by modifying the analysis.  The
quality of the fit was checked by construction of the pull-off
diagrams.

We find that the LMA solution with parameters in the range $\Delta m^2
\sim (4 - 5) \cdot 10^{-5}$ eV$^2$ and $\tan^2 \theta = 0.35 - 0.40$
gives the best fit to the data. Moreover, the solution reproduces the
flat spectrum of recoil electrons in SK and gives very good
description of the rates and of the Day-Night asymmetry.  It is in a
very good agreement with SSM fluxes of the boron ($f_B = 1.10 - 1.13$)
and $hep-$neutrinos.  The values of the oscillation parameters and the
goodness of the fit are stable with respect to variations of the
analysis.

The LOW solution appears at 90 \% C.L. with respect to 
the best fit point in the LMA range. It gives  good fit to
the SK spectral data, but rather poor fit to the rates: it 
predicts larger than measured $Ar-$production rate and smaller than
measured $Ge-$production rate.

The VAC solution has high goodness of the global fit 
due to a very good description of the day and the night spectra measured
by SK.  It has, however, problems with other criteria.  The solution
requires strong deviation of the boron and $hep-$fluxes from their
SSM values. The fit becomes substantially
worse when SSM restrictions are applied. The solution shows strong
deviations in the pull-off diagram (especially for the SuperKamiokande
rate and the $Ar-$production rate).  The analysis of the rates only
selects a different point in the oscillation parameter space.

The SMA appears at about $3\sigma$ C.L. level with respect to the best
global solution in the LMA range. It gives bad fit of the recoil
electron energy spectrum and also the fit of the rates is rather
poor.

The best sterile solution is VAC(sterile) with $\Delta m^2 = 1.4
\cdot 10^{-10}$ eV$^2$.  It is the only sterile solution accepted at
lower than $3 \sigma$ C.L. in the free flux analysis. This solution, 
however,  has serious problems with other tests.  Similarly to
VAC(active) the solution requires strong deviation of the boron and
$hep-$ fluxes from their SSM values. The
fit becomes substantially worse when SSM restrictions are applied. The
solution shows strong deviations in the pull-off diagram: in
particular, the $\nu e-$ event rate is about $3 \sigma$ below the SK
rate, the $Ar-$production rate is   $2.4 \sigma$ higher than the
Homestake result. The analysis of the rate only selects a different
point in the oscillation parameter space. Significant distortion of
the energy spectrum is expected. 
Appearance  of the VAC solutions (both active and sterile) 
in the global fit is related to a small spread  of the experimental points 
in the SK spectrum.  These solutions can describe rather precisely 
the bump at 7 - 8 MeV and the dip at 11 - 12 MeV.

Maximal mixing is allowed at $3\sigma$ level in the LMA region and 
at 99 \% C.L.  in the LOW and quasivacuum oscillation regions.

Measurements of the D-N asymmetry will provide strong discrimination
among the solutions. In the best fit range of the LMA solution the
asymmetry in SNO is $A_{DN}^{SNO} \simeq (7 - 8) \%$, and in the
$3\sigma$ allowed region it can reach 15 - 20 \%.  In the LOW region
one expects smaller asymmetry: in the best fit point $A_{DN}^{SNO}
\simeq (2 - 3)\%$,  although in the $3\sigma$ allowed region it can 
be as large  as 10 - 12 \%. 
For the SMA solution the asymmetry is $A_{DN}^{SNO}  \leq 3 \%$.  
The predicted SMA zenith angle distribution is not supported by SK data.

Clearly, observation of a large D-N asymmetry $A_{DN}^{SNO} > 5 \%$
will favor the LMA solution. It will strongly disfavor the SMA
solution, and exclude solutions based on 
vacuum oscillations (VAC, QVO and and Just-so$^2$).  Furthermore,
this result will strongly restrict LMA to lower $\Delta m^2$. 
Even with existing statistics (expected error is 3 - 4 \%) any SNO
result on the D-N asymmetry will be of  physical importance
excluding some solutions or further restricting the allowed regions.

Significant distortion of the boron neutrino spectrum is expected for
SMA and VAC solutions and already the present data can affect them.
The VAC solution predicts the bump at 6 MeV and the dip at 11 MeV   
in the dependence of the $R_{CC}$ on energy.  
In the case of LMA solution turn up of the spectrum at low energies is
expected which  will be rather difficult to observe  with
SNO. For LMA and LOW one expects practically no distortion in future SNO
measurements.

Important discrimination among solutions can be done  using precise  
determination of the  [NC]/[CC] ratio. 

With forthcoming SNO data, we have a chance to make further major step in
identification of the solution of the solar neutrino problem.

\section*{Acknowledgments}

The authors are grateful to J. Beacom, E. Lisi and M. C. Gonzalez-Garcia 
for  comments on the first version of the paper. 
We thank  Y. Suzuki for clarification 
of the way the SuperKamiokande collaboration treat  
the hep neutrino flux in the analysis of the data. 
This work was supported by DGICYT under grant PB95-1077 and by the TMR
network grant ERBFMRXCT960090 of the European Union. 
The calculations were performed on the computers at the Institute
for Advanced Study in Princeton and the Nuclear Astrophysics Group at
the Department of Physics and Astronomy, UW-Madison.  Support for the
calculations has been provided under NSF grants No.PHY0070928 and
No.9605140.

\newpage

\newpage
\vskip -2cm
\begin{figure}[ht]
\centering\leavevmode
\epsfxsize=0.9\hsize
\epsfbox{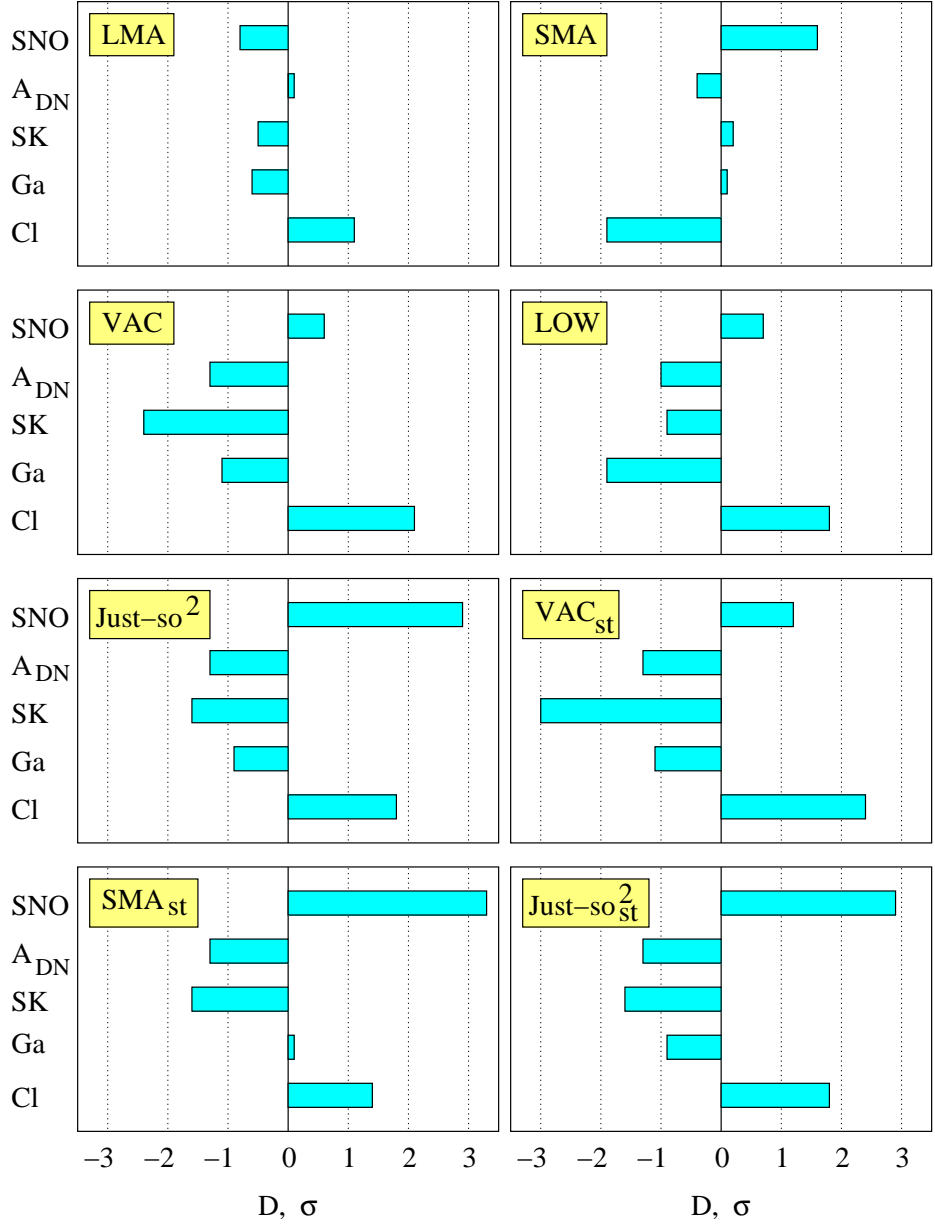}
\vskip -1.5cm
\caption{
Pull-off diagrams for global solutions. Shown are deviations 
of predictions for the $Ar-$production rate  
$Ge-$production rate, SK rate, the day-night asymmetry at SK, 
and the SNO rate  
from experimentally measured values. 
The pull-offs  are expressed  in the units of 1 standard deviation,  
$1 \sigma$.}
\label{pull} 
\end{figure}

\begin{figure}[ht]
\centering\leavevmode
\epsfxsize=1.0\hsize
\epsfbox{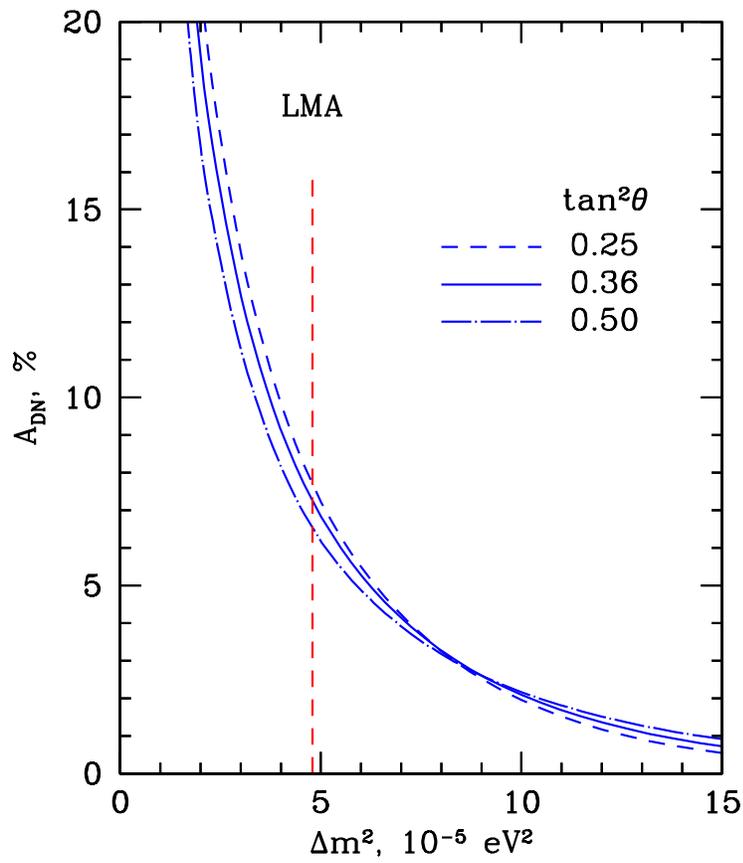}
\vskip -2cm 
\caption{
The dependence of the Day-Night asymmetry of the $CC-$event rate
measured at SNO on the $\Delta m^2$ for different values of the mixing
angle. The oscillation parameters are taken from the LMA allowed region.
The best fit  value  of $\Delta m^2$ is marked by the vertical dotted
line.}
\label{dn-l} 
\end{figure}


\begin{figure}[ht]
\centering\leavevmode
\epsfxsize=1.0\hsize
\epsfbox{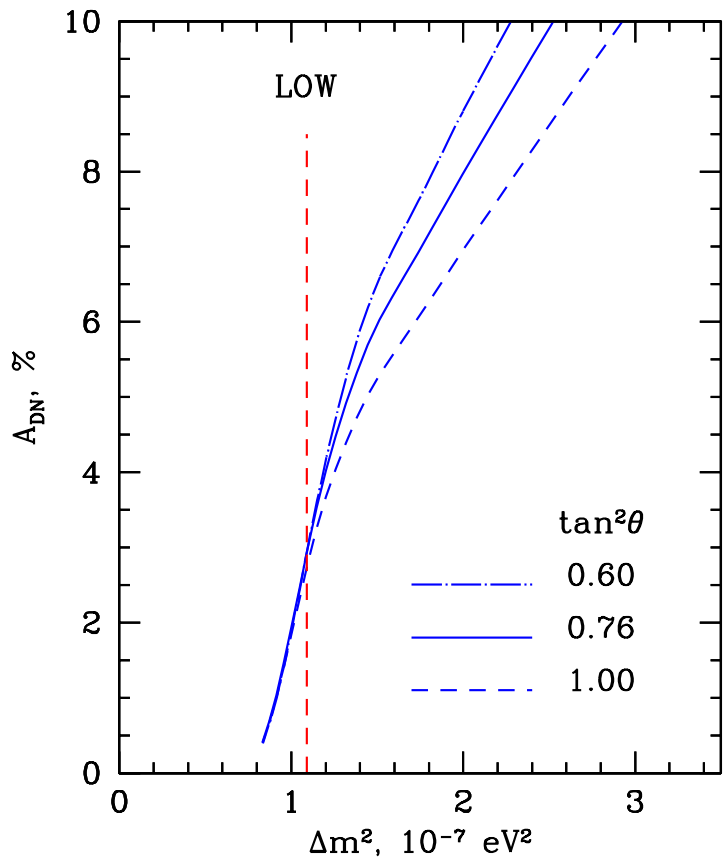}
\vskip -2cm
\caption{
The same as in the Fig. 3a for the LOW solution. 
The dotted vertical lines mark the best fit values of 
$\Delta m^2$ from the free flux fit (left) and SSM restricted fit 
(right).   
}
\label{dn-low} 
\end{figure}

\begin{figure}[ht]
\centering\leavevmode
\epsfxsize=1.0\hsize
\epsfbox{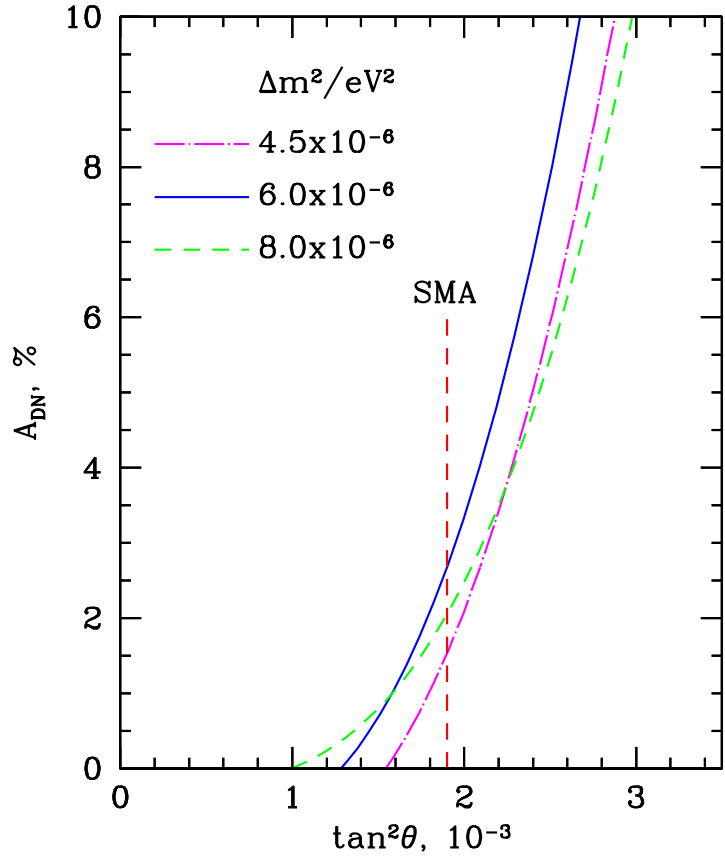}
\vskip -2cm
\caption{
The dependence of the Day-Night asymmetry of the $CC-$event rate
measured at SNO on $\tan^2 \theta$
for different values of $\Delta m^2$.
The oscillation parameters are taken from the SMA allowed region.
The best fit  value of $\tan^2 \theta$  is marked 
by the vertical dotted line.
}
\label{dn-sma} 
\end{figure}


\begin{figure}[ht]
\centering\leavevmode
\epsfxsize=1.1\hsize
\epsfbox{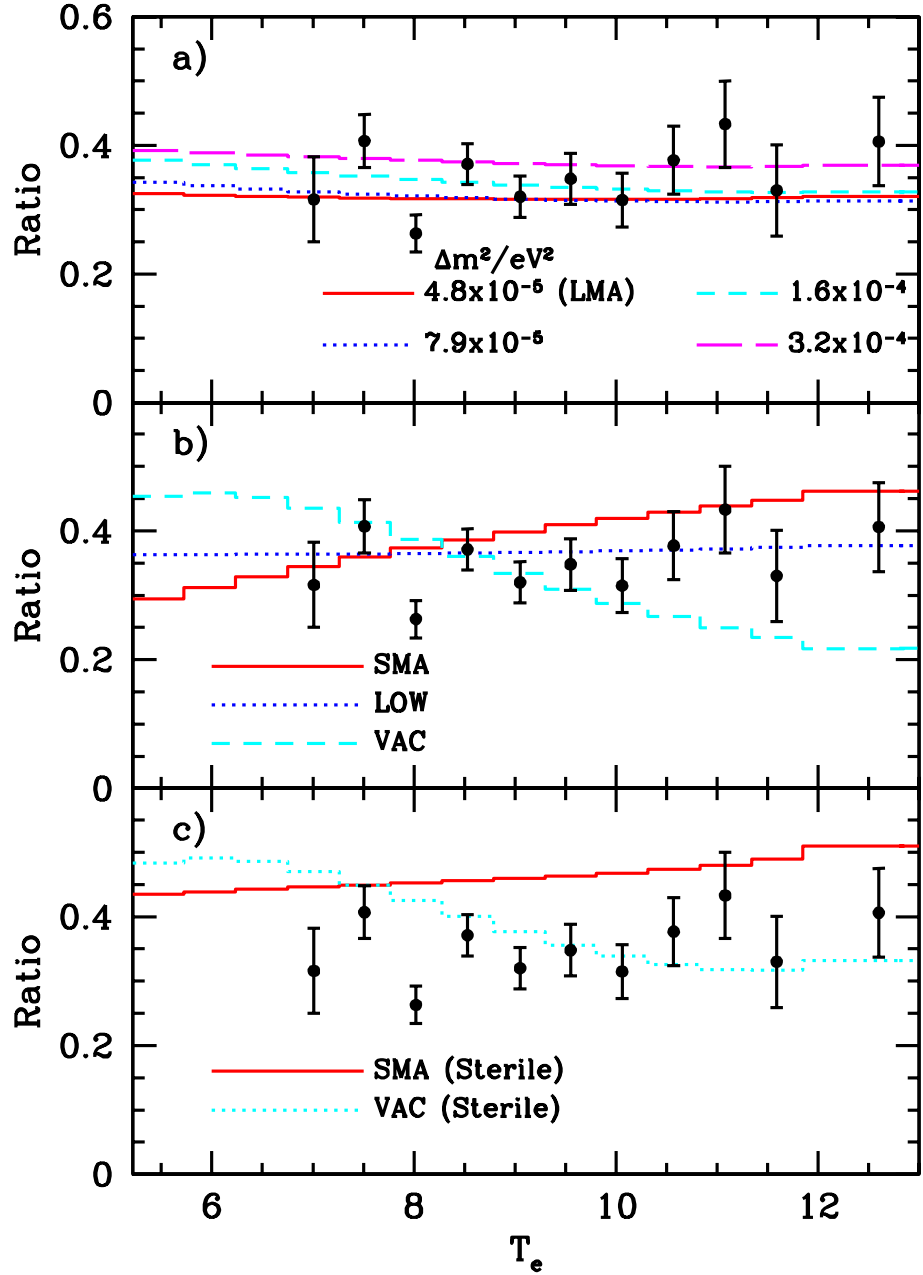}
\caption{
The recoil electron energy spectra of the $CC-$events in  SNO for 
global solutions of the solar neutrino problem. 
Shown are the ratio of the 
number of events with and without conversion
as a function of the electron kinetic energy.  
a). The spectra in  the LMA solution region     
for different values of
$\Delta m^2$ and $\tan^2 \theta = 0.35$.
b).  Spectra for the best fit points of the SMA, LOW and VAC 
(active) solutions. 
c). Spectra for the best fit points of the SMA(sterile) and 
VAC(sterile) solutions.  
Shown also in all panels is  the SNO experimental data. 
}
\label{spectra} 
\end{figure}


\begin{figure}[ht]
\centering\leavevmode  
\epsfxsize=0.9\hsize
\epsfbox{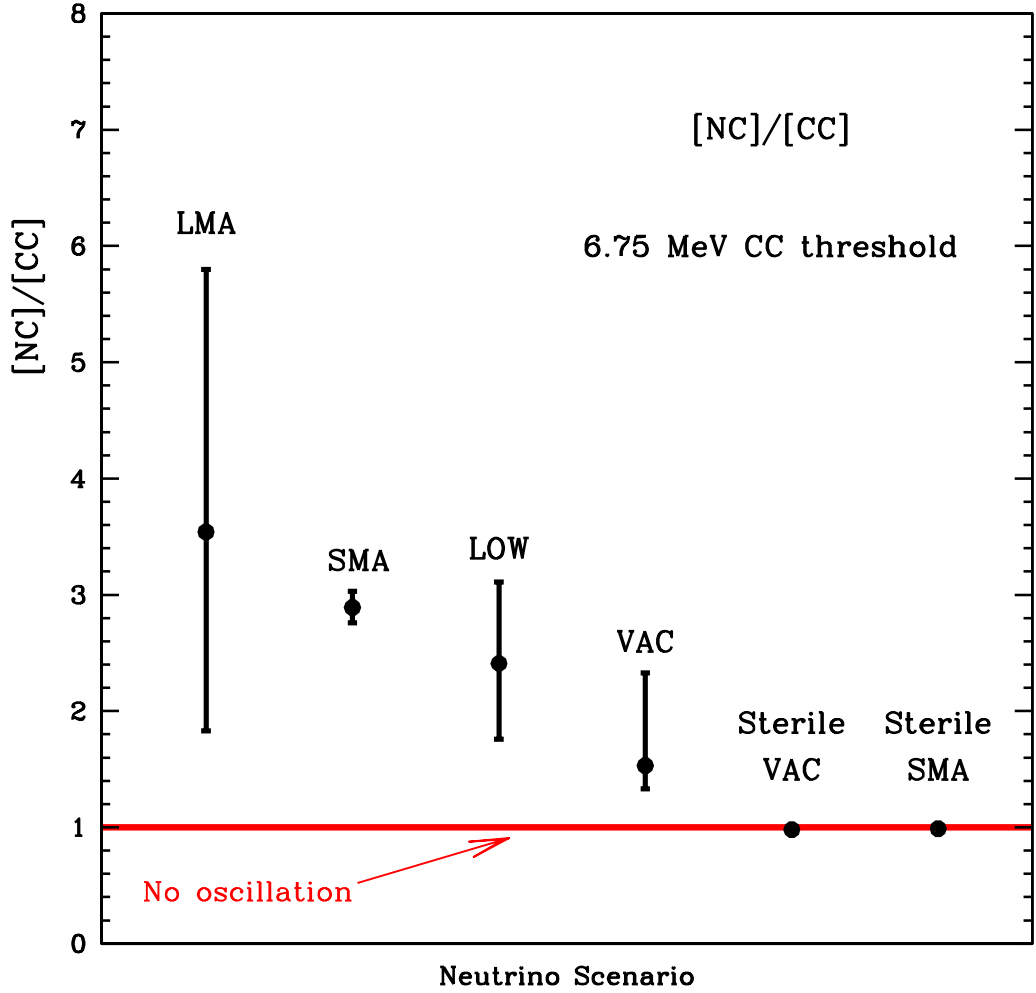}
\caption{The ratio  of the  reduced NC - event rate 
to the reduced  CC- event rate. 
The circles give  
values of  [NC]/[CC] in the best fit points of the global solutions. 
The error bars show the prediction intervals of  [NC]/[CC]  
which correspond to the $3\sigma$ allowed regions of global
solutions. 
}
\label{nc-cc}
\end{figure}

\begin{figure}[ht]
\centering\leavevmode  
\epsfxsize=0.9\hsize
\epsfbox{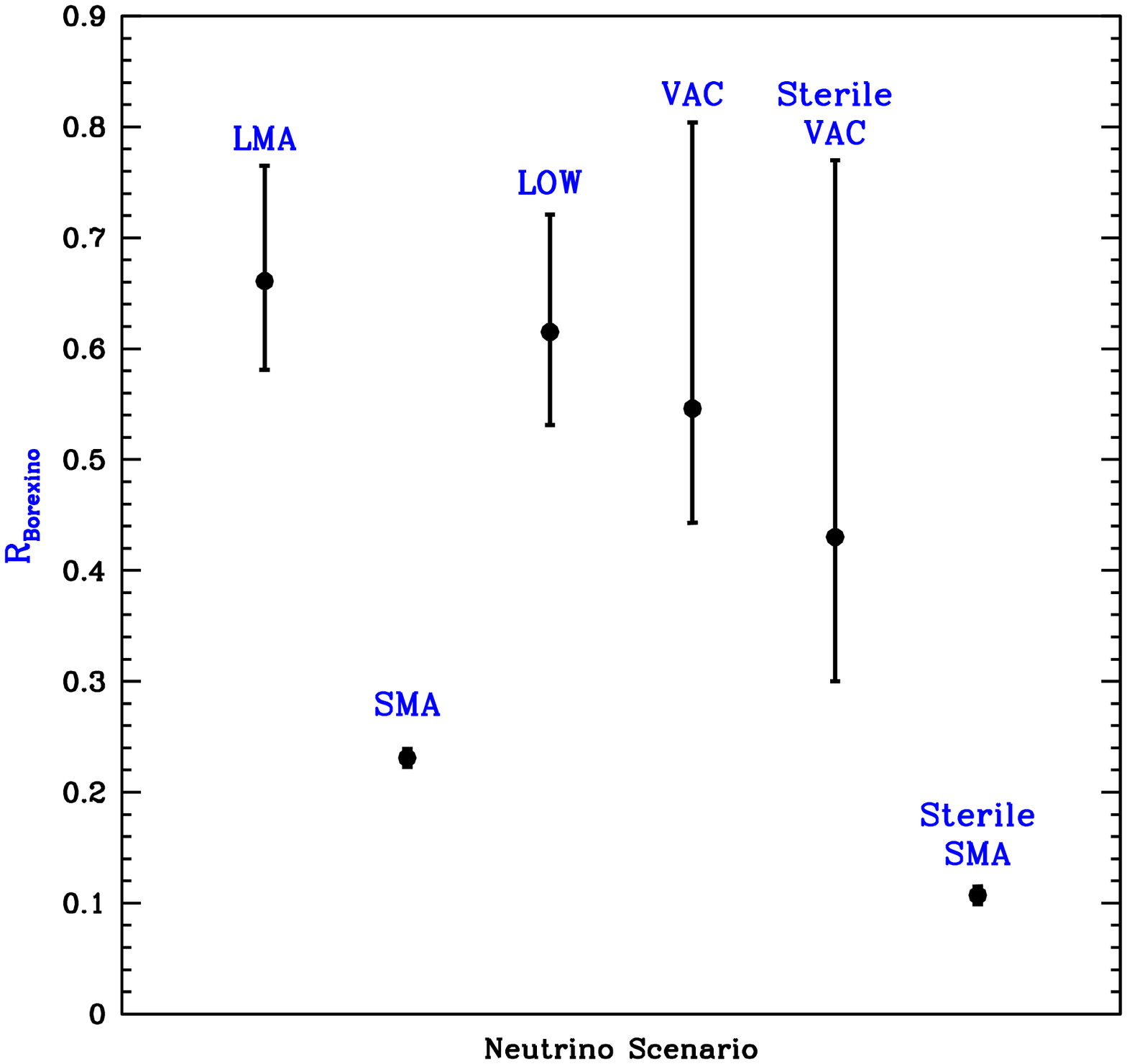}
\caption{
The reduced event rate in the BOREXINO experiment. 
The circles give
values of $R_{borexino}$ in the best fit points of the global solutions.
The error bars show the prediction intervals 
which correspond to the $3\sigma$ allowed regions of global
solutions.   
}
\label{borex}
\end{figure}


\begin{thebibliography}{99}



\bibitem{sno} Q. R. Ahmad {\it et al.}, SNO collaboration, nucl-ex/0106015. 
 
\bibitem{sno1} A. B. McDonald,  Proc. of the 19th Int. Conf. on Neutrino 
Physics and Astrophysics,  {\it Neutrino 2000} , Sudbury, Canada 2000, 
Nucl. Phys. B (Proc. Suppl.) {\bf 91} (2001) 21.  


\bibitem{SK} S. Fukuda {\it et al.} (SuperKamiokande collaboration) 
hep-ex/0103032.  


\bibitem{astro} J. N. Bahcall, hep-ph/0108147, hep-ph/0108148; 
V. Berezinsky hep-ph/0108166.


\bibitem{bks01} J. N. Bahcall, P.I. Krastev, A. Yu. Smirnov, JHEP {\bf 5} 
(2001) 15.

\bibitem{sno-sk} See previous analysis of this type:  
Waikwok Kwong, S. P. Rosen, Phys. Rev. {\bf D 54} (1996) 2043;  
F.L. Villante, G. Fiorentini, E. Lisi, Phys. Rev. 
{\bf D59} (1999)  013006; 
 
J. N. Bahcall, P.I. Krastev, A. Yu. Smirnov, Phys. Lett. {\bf B 477} (2000) 
401. 



\bibitem{barger} V. Barger, D. Marfatia and K. Whisnant, hep-ph/0106207. 

\bibitem{fogli} G.L. Fogli, E. Lisi, D. Montanino and A. Palazzo, 
hep-ph/0106247. 

\bibitem{valencia} J. N. Bahcall, M. C. Gonzalez-Garcia, Carlos Pena-Garay, 
hep-ph/0106258.

\bibitem{calcutta} A. Bandyopadhyay, S Choubey, S. Goswami, K. Kar, 
hep-ph/0106264.

\bibitem{gouvea} A. de Gouvea and C. Pena-Garay, hep-ph/0107186. 

\bibitem{BS} R. Barbieri, A. Strumia, hep-ph/0011307,  v4 (July 2001).

\bibitem{giunti} C. Giunti, hep-ph/0107310.

\bibitem{berezinsky}  V. Berezinsky  M. Lissia,  hep-ph/0108108. 


\bibitem{bks98}J.N. Bahcall, P.I. Krastev and A.Yu. Smirnov,
Phys. Rev. {\bf D58} 096016 (1998).

\bibitem{bks-10} J. N. Bahcall, P.I. Krastev, A. Yu. Smirnov, Phys. Rev. 
{\bf D 62}  (2000) 093004.

\bibitem{bks-corr} J. N. Bahcall, P.I. Krastev, A. Yu. Smirnov, Phys. Rev. 
{\bf D 63} (2001) 053012. 


\bibitem{Cl} B. T. Cleveland  {\it et al} Astroph. J. {\bf 496} (1998) 505; 
K. Lande {\it et al}, in  {\it Neutrino 2000} \cite{sno1}, p.50. 

\bibitem{sage} V. Gavrin, (SAGE collaboration) 
Proc. of the 19th Int. Conference
 on Neutrino Physics and Astrophysics,  {\it Neutrino 2000} , Sudbury, 
Canada, June 2000,  Nucl. Phys. B (Proc. Suppl.) {\bf 91} (2001) 36.  


\bibitem{gallex} W. Hampel {\it et al.} (GALLEX Collaboration) Phys. Lett. 
{\bf B 447} (1999) 127. 

\bibitem{gno} M. Altmann {\it et al.} (GNO Collaboration) Phys. Lett. 
{\it B 490} (2000) 16; 
E. Bellotti {\it et al.} Proc. of the XIX Int. Conference 
on Neutrino Physics and Astrophysics,  {\it Neutrino 2000}, Sudbury, 
Canada 2000,  Nucl. Phys. B (Proc. Suppl.) {\bf 91} (2001) 44.  


\bibitem{ssm} J. N. Bahcall, M.H. Pinsonneault and S. Basu,  
 Astrophys. J. {\bf 555} (2001)990.  


\bibitem{KSm} P. I. Krastev and A. Yu. Smirnov, Phys. Lett. 
{\bf B338} (1994) 282. 

\bibitem{hepfl} 
L. E. Marcucci {\it et al.}, Phys. Rev. {\bf C63} (2001)  015801;   
T.-S. Park, {\it et al.}, hep-ph/0107012 and references 
therein. 



\bibitem{css} P. Creminelli, G. Signorelli, A. Strumia, JHEP {\bf 0105} 
(2001) 052,  hep-ph/0103179.


\bibitem{lma99} J. N. Bahcall, P.I. Krastev, A. Yu. Smirnov, Phys. Rev. 
{\bf D60} (1999) 093001. 

\bibitem{nufac} S. Geer, Phys. Rev. {\bf D57} (1998) 6989.  


\bibitem{CHOOZ} CHOOZ Collaboration, M. Apollonio {\it et al.},
Phys.Lett. {\bf B420}, 397 (1998).


\bibitem{core-bin} 
S. P. Mikheyev and A. Yu. Smirnov, {\it '86 Massive Neutrinos 
in Astrophysics and in Particle Physics} Proc. of the 6th Moriond
Workshop, edit by O. Fackler and J. Tran Thanh Van (Edition Frontiers 
Gif-sur-Yvette, 1986) p. 355;  
A. J. Baltz and J. Weneser, Phys. Rev. {\bf D50}  5971 (1994), 
{\it ibid } {\bf D51} (1994) 3960; 
E. Lisi, D. Montanino, Phys. Rev. {\bf D56} (1997) 1792; 
J. M. Gelb, Wai-kwok Kwong, S. P. Rosen, Phys. Rev. Lett. 
{\bf 78} (1997) 2296.     

\bibitem{par-res}
S. T. Petcov, Phys. Lett. {\bf B434} (1998); 
E. Kh. Akhmedov, Nucl. Phys.  {B 538} (1999) 25; 
M. V. Chizhov and S. T. Petcov, Phys. Rev. Lett. 
{\bf 83} (1999) 1096.  

\bibitem{bls} K. S. Babu, Q. Y. Liu, A. Yu. Smirnov, Phys. Rev. {\bf D57}  
(1998) 5825. 


\bibitem{max} M. C. Gonzalez-Garcia, C. Pena-Garay, Y. Nir,
A. Yu. Smirnov, 
Phys. Rev. {\bf D63} (2001) 013007. 


\bibitem{dvali} G. Dvali and A. Yu. Smirnov,  
Nucl. Phys. {\bf B563} (1999)  63. 

\bibitem{bulk} A. Lukas, P. Ramond, A. Romanino, G. G. Ross,  
JHEP 0104:010, 2001;  D.O. Caldwell, R.N. Mohapatra,
S.J. Yellin, hep-ph/0102279.  

\bibitem{B-K} J. N. Bahcall and P. I. Krastev, Phys. Rev.  {\bf C56} (1997) 
2839. 

\bibitem{M-P} M. Maris and S. T. Petcov, Phys. Rev. {\bf D62} (2000) 093006. 

\bibitem{flmp} G. L. Fogli, E. Lisi, D. Montanino, A. Palazzo,  
hep-ph/0008012. 

\bibitem{zenith} M. C. Gonzalez-Garcia,  C. Pena-Garay,  A. Yu. Smirnov, 
Phys. Rev. {\bf D63} (2001) 113004.




\bibitem{bl-mom} J. N. Bahcall and E. Lisi, Phys. Rev. {\bf D54} (1996) 
5417.  

\bibitem{bkl} J. N. Bahcall, P. I. Krastev and E. Lisi, 
Phys. Rev. {\bf C55} (1997) 494. 


\bibitem{flm-mom} G. L. Fogli, E. Lisi and D. Montanino, Astropart. Phys. 
{\bf 9} (1998) 119. 

\bibitem{bor} BOREXINO Collaboration, G. Ranucci et al., 
19th International Conference on Neutrino Physics and
Astrophysics - Neutrino 2000, Sudbury, Ontario, Canada, 16-21 June 2000, 
Nucl. Phys. Proc. Suppl. {\bf 91} (2001)  58.  

\bibitem{BKSirlin} J. N. Bahcall,
M. Kamionkowski, A. Sirlin, Phys. Rev. {\bf D51} 
(1995) 6146.  

\bibitem{bor-dn} A.J. Baltz,  J. Weneser, 
Phys. Rev. {\bf D50} (1994) 5971,  Addendum-ibid. {\bf D51} (1995) 3960;   
for recent analysis see  A. de Gouvea, A. Friedland, H. Murayama,  
JHEP {\bf 0103} (2001) 009;   
G.L. Fogli, E. Lisi, D. Montanino,
A. Palazzo,  Phys. Rev. {\bf D61}  (2000)  073009. 

\bibitem{bor-seas} For recent analysis see  
B. Faid, G.L. Fogli, E. Lisi, D. Montanino,  
Astropart. Phys., {\bf 10}  (1999) 93;  
A. de Gouvea, A. Friedland, H. Murayama,  
Phys. Rev. {\bf D60}  (1999) 093011.  



\end{thebibliography}
\end{document}